\PassOptionsToPackage{table}{xcolor}
\documentclass[fleqn,10pt]{wlscirep}
\usepackage[utf8]{inputenc}
\usepackage[T1]{fontenc}

\usepackage{times}  
\usepackage{helvet}  
\usepackage{courier}  
\usepackage{subcaption}
\usepackage{multirow}
\usepackage{ulem}
\usepackage{graphicx}

\usepackage{lineno}

\let\oldtextsf\textsf

\renewcommand{\textsf}[1]{{\fontsize{8.5}{10}\selectfont\oldtextsf{#1}}}

\newcommand{\textsfsm}[1]{{\fontsize{5}{5}\selectfont\oldtextsf{#1}}}

\title{A Novel Fusion Architecture for PD Detection Using Semi-Supervised Speech Embeddings}

\author[1,$\dagger$,*]{Tariq Adnan}
\author[1,$\dagger$]{Abdelrahman Abdelkader}
\author[1,$\ddagger$]{Zipei Liu}
\author[1,$\ddagger$]{Ekram Hossain}
\author[1]{Sooyong Park}
\author[1,+]{Md Saiful Islam}
\author[1,+]{Ehsan Hoque}

\affil[1]{Department of Computer Science, University of Rochester, Rochester, New York, USA}
\affil[*]{tadnan@ur.rochester.edu}
\affil[$\dagger$]{These authors contributed equally to this work.}
\affil[$\ddagger$]{These authors also contributed significantly and equally between them.}
\affil[+]{These authors jointly supervised this work.}

\begin{abstract}
We introduce a framework for screening Parkinson's disease (PD) using English pangram utterances. Our dataset includes $1,306$ participants ($392$ with PD) from both home and clinical settings, covering diverse demographics ($53.2\%$ female). We used deep learning embeddings from Wav2Vec 2.0, WavLM, and ImageBind to capture speech dynamics indicative of PD. Our novel fusion model for PD classification aligns different speech embeddings into a cohesive feature space, outperforming baseline alternatives. In a stratified randomized split, the model achieved an AUROC of $88.9\%$ and an accuracy of $85.7\%$. Statistical bias analysis showed equitable performance across sex, ethnicity, and age subgroups, with robustness across various disease durations. Detailed error analysis revealed higher misclassification rates in specific age ranges for males and females, aligning with clinical insights. External testing yielded AUROCs of $82.1\%$ and $78.4\%$ on two clinical datasets, and an AUROC of $77.4\%$ on an unseen general spontaneous English speech dataset, demonstrating versatility in natural speech analysis and potential for global accessibility and health equity.
\end{abstract}
\begin{document}

\flushbottom
\maketitle
%
%


\section{Introduction}
The diagnosis of Parkinson's Disease (PD) is traditionally reliant on the clinical assessments focused on the motor symptoms of the individuals~\cite{goetz2008movement}.
Traditional methods, while effective, often miss the subtle early symptoms of the disease, leading to delayed interventions ~\cite{morel2022patient}. 
The situation is further exacerbated by the limited accessibility to specialized neurological healthcare, particularly in regions with significantly lower ratios of neurologists to the population. 
For instance, Bangladesh had only 86 neurologists for over 140 million people in 2014~\cite{chowdhury2014pattern}, while some African nations had one neurologist per three million people, with 21 countries having fewer than five neurologists each~\cite{kissani2022does}.
Given the expected doubling of PD cases by 2030~\cite{dorsey2007projected}, there's a pressing need for accessible, home-based diagnostic solutions to address global disparities in healthcare access.

Recent advancements have seen a shift towards integrating digital biomarkers to develop automated AI based at-home PD detection and progression tracking tools~\cite{rahman2021detecting,dorsey2016moving,dubey2015echowear,yang2022artificial,islam2023using}.
Techniques vary from sensor-based nocturnal breathing signal~\cite{yang2022artificial} and accelerometric data collection~\cite{schalkamp2023wearable}, to digital analysis of facial expressions~\cite{adnan2023unmasking}.
However, wearables and sensors may be inconvenient for the elderly~\cite{wang2019alignment,forkan2019internet}, and posed expressions can miss subtle diagnostic cues. 

Alternatively, speech analysis offers a non-invasive route, leveraging natural speech patterns for PD detection.
Traditional speech analysis in PD --- primarily relied on sustained phonation tasks~\cite{little2007exploiting,rahman2021detecting,saldanha2023jitter,hawi2022automatic,upadhya2017statistical,canturk2016machine,rusz2011quantitative,liu2012vocal} --- although useful, does not reflect the complexities of natural speech.
To counter that, studies have been proposed to use continuous speech to develop PD classifiers using varying technologies such as CNNs~\cite{frid2016diagnosis}, time-frequency analysis~\cite{umapathy2005discrimination}, or SVMs~\cite{khan2014classification}. However, these studies rely on fixed recording setups and small sample sizes, which limits the generalizability of the models and fail to adequately address accessibility concerns.
{Even with larger datasets, models like CNNs and SVMs face structural limitations. CNNs, while powerful for feature extraction, are primarily designed for spatial data, and their application to time-series or speech data can be challenging unless properly adapted~\cite{bai2018empirical,ravanelli2018speaker}. They may require deep architectures to capture complex temporal dependencies in PD speech data, increasing the risk of overfitting if the model is not properly regularized or if the dataset lacks sufficient variability to cover real-world scenarios~\cite{szegedy2017inception}. SVMs, although effective on small, well-separable data, may struggle in high-dimensional, noisy feature space~\cite{ben2008support} common in speech signals~\cite{zeng2007survey} due to their limitation in temporal pattern handling compared to neural networks~\cite{graves2013speech}. These structural lacking of the modeling architectures in the existing literature underscore the demand of more adaptable models, which can capture the nuanced vocal changes associated with PD while generalizing better across diverse recording conditions and larger datasets.}


{Traditional feature engineering, while interpretable, requires extensive time, human effort, and domain expertise to identify and select relevant features manually~\cite{guyon2003introduction}.}
{This approach introduces the risk of human bias, as researchers may prioritize familiar or well-documented features, potentially overlooking subtle yet critical nuances present in neurological speech patterns.
Studies have shown that even with careful feature selection, critical variations indicating early-stage PD can be missed, reinforcing the need for more automated, data-driven feature extraction techniques, such as those used in deep learning models~\cite{paul2022bias, rashnu2024integrative, wang2020robust, jhapate2024gait}.} 
{Recent advancements in semi-supervised learning models, such as WavLM~\cite{chen2022wavlm}, Wav2Vec 2.0~\cite{baevski2020wav2vec}, and ImageBind~\cite{girdhar2023imagebind}, offer pre-trained, openly accessible models that capture complex and abstract representations of speech data.}
{These models, trained on diverse, large-scale datasets, exhibit strong performance in several downstream applications like automatic speech recognition (ASR)\cite{baevski2020wav2vec,hsu2021hubert} and speech diarization\cite{chen2022wavlm}, suggesting their embeddings capture intricate, high-dimensional speech dynamics.}
{
These embeddings hold significant promise to reveal subtle voice changes associated with the PD that handcrafted features might miss.}
Despite their potential, only limited works~\cite{klempivr2023evaluating} have been done to explore semi-supervised acoustic models' utility in PD classification. 
{Although this studys~\cite{klempivr2023evaluating} was conducted only on a small cohort of 60 participants (28 PD), it found significant promise in applying semi-supervised speech embeddings from wav2vec in PD classification for both English and Italian language.}
On the other hand, fusion after projection of one feature set into another latent space can achieve significant improvement in feature alignment, noise reduction, and dimensionality consistency~\cite{lin2021efficient,veli2022transformer,wang2019alignment}, simplifying the architecture and enhancing cross-modal learning capacity~\cite{lu2019vilbert,chen2020uniter,li2019visualbert}. 
Despite the significant advancements in speech processing and disease classification, the literature has yet to fully explore the potential of combining semi-supervised deep speech embeddings with projection-based fusion models for PD classification. 

\begin{figure*}[t]
    \centering
    \hspace*{-7mm}
    \includegraphics[width=1.05\linewidth]{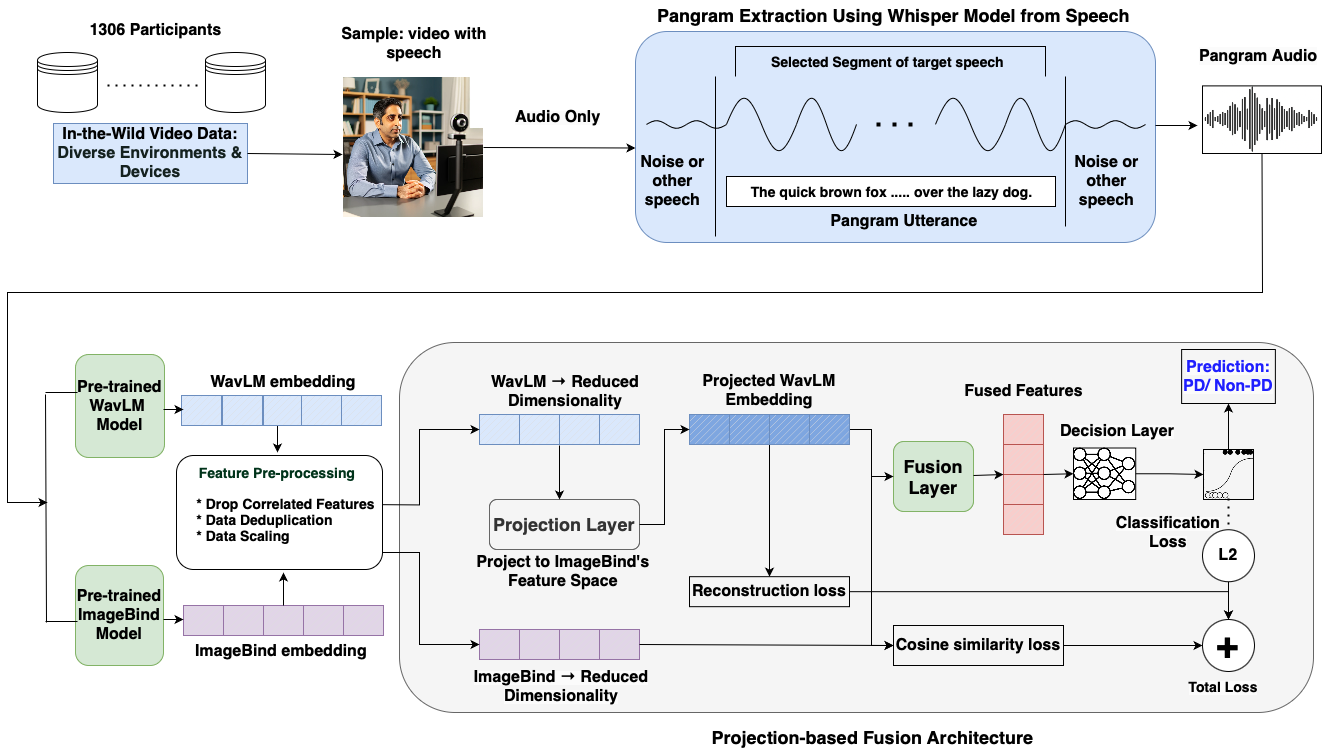}
    \caption{\textbf{Our proposed framework of fusion based PD classifier using deep embeddings from WavLM and ImageBind.} First, the speech is separated from video datasets. Then the segment of the audio file where the participants utter the pangram is separated. Vector embeddings from the last layers of WavLM and ImagBind are extracted for the speech data. Then WavLM feautures are projected into the space of ImagBind features set. Finally the projected features are fused and passed through a classification layer that can determine the participant as PD or control. Note that the image of the person is AI generated.}
    \label{fig:overview}
\end{figure*}

{
In our study, we leveraged a large-scale dataset of $1,306$ participants, including $392$ PD-diagnosed individuals, collected from varied environments --- participants' homes, clinical settings, and PD care centers --- enabling us to overcome the limitations of smaller cohorts and constrained recording environments in PD detection. 
We expanded the application of semi-supervised vector embeddings of free-flow speech through a novel projection-based fusion approach, demonstrating that fusing deep embeddings from WavLM to ImageBind feature space enhances the model's ability to capture PD-specific speech characteristics.
}
This fusion approach notably outperformed traditional concatenation by effectively aligning embeddings, reducing redundancy, and enabling a more nuanced representation of disease indicators within speech.
To prepare the dataset, participants were instructed to record themselves while uttering the English pangram that starts with ``the quick brown fox.''
We modeled the free-flow speech using the pangram utterance to simplify the analysis as it ensures a consistent speech content across all participants.
{This study highlights that semi-supervised embeddings are highly effective for PD classification when paired with projection-based fusion, marking a key contribution in identifying PD-specific vocal nuances often missed by handcrafted features.}
{We maximized the generalizability and robustness of the model by collecting data across diverse environments --- homes, clinics, and a PD care facility --- and a demographically diverse cohort.
}
We conducted extensive error analysis using multiple methodologies to identify any comparatively underperforming demographic subgroups.
{With our web-based framework, English-speaking individuals having access to a webcam-enabled laptop or desktop can record their speech and receive a PD screening. This approach can enhance accessibility, particularly in regions with limited access to neurological care.}
Figure~\ref{fig:overview} illustrates our PD classification pipeline, which processes raw video inputs to determine PD presence effectively.

\section{Results}

\subsection{Dataset}
We collected our dataset from $1306$ participants, comprising $392$ of them diagnosed with PD and $914$ without the condition. We used PARK~\cite{langevin2019park}, a web-based framework for data collection purpose. Using the framework,
each participant recorded themselves in front of the web camera while reciting the ``quick brown fox" English pangram
across three distinct recording environments -- \textsf{Home Recorded}, \textsf{Clinical Setup}, and \textsf{PD Care Facility} -- each varying in terms of ambient noise and data collection setting and/or equipment. At the \textsf{Clinical Setup}, and \textsf{PD Care Facility}, some participants provided multiple data samples and eventually, we obtained $1,854$ video clips having audio of pangram utterance. Note that PD labels of the participants from \textsf{Clinical Setup} and \textsf{PD Care Facility} cohorts are clinically validated, while the \textsf{Home Recorded} labels are self-reported.
The demographic information of the participating cohort is detailed in Table~\ref{tab:demographics}.

\renewcommand{\arraystretch}{1.2}

\begin{table}[!htb]
\centering
\caption{\textbf{Demographic information of the participants.}}
\resizebox{0.8\columnwidth}{!}{
\begin{tabular}{cclrrr}
\toprule

\multicolumn{1}{l}{\textbf{Demographic Properties}} && \multicolumn{1}{l}{\textbf{Attribute}} & \multicolumn{1}{c}{\textbf{With PD}} & \multicolumn{1}{c}{\textbf{Without PD}} & \multicolumn{1}{c}{\textbf{Total}} \\ \midrule

&&\multicolumn{1}{l}{Number of Participants} & \textbf{392} \textbf{(30.0\%)} & \textbf{914} \textbf{(70.0\%)} & \textbf{1306} \textbf{(100.0\%)} \\ \midrule

\multirow{4}{*}{\textbf{Sex}} 
  && Female & 171 (43.6\%) & 524 (57.3 \%) & \textbf{695} \textbf{(53.2\%)} \\ 
  && Male & 218 (55.6\%) & 390 (42.7 \%) & \textbf{608} \textbf{(46.6\%)} \\ 
  && Nonbinary & 1 (0.3\%) & 0 (0 \%) & \textbf{1} \textbf{(0.1\%)} \\ 
  && Unknown & 2 (0.5\%) & 0 (0 \%) & \textbf{2} \textbf{(0.2\%)} \\
\midrule

 && Below 20 & 0 (0.0\%) & 8 (0.9\%) & \textbf{8 (0.61\%)} \\
 && $20-29$ & 2 (0.5\%) & 41 (4.5\%) & \textbf{43 (3.29\%)} \\
 && $30-39$ & 3 (0.8\%) & 24 (2.6\%) & \textbf{27 (2.07\%)} \\
 \textbf{Age in years} 
 && $40-49$ & 9 (2.3\%) & 20 (2.2\%) & \textbf{29 (2.22\%)} \\
 range: $16.0-93.0$ 
 && $50-59$ & 40 (10.2\%) & 160 (17.5\%) & \textbf{200 (15.31\%)} \\
 mean: $62.2$, std: $12.7$ 
 && $60-69$ & 122 (31.1\%) & 349 (38.2\%) & \textbf{471 (36.06\%)} \\
 && $70-79$ & 142 (36.2\%) & 107 (11.7\%) & \textbf{249 (19.07\%)} \\
 && 80 and above & 23 (5.9\%) & 7 (0.8\%) & \textbf{30 (2.3\%)} \\
 && Unknown & 51 (13.0\%) & 198 (21.7\%) & \textbf{249 (19.07\%)} \\
\midrule

\multirow{6}{*}{\textbf{Ethnicity}}
 && White & 223 (56.9\%) & 638 (69.8\%) & \textbf{861 (65.9\%)} \\
 && Black or African American & 6 (1.5\%) & 42 (4.6\%) & \textbf{48 (3.7\%)} \\
 && American Indian or Alaska Native & 1 (0.3\%) & 4 (0.4\%) & \textbf{5 (0.4\%)} \\
 && Asian & 4 (1.0\%) & 55 (6.0\%) & \textbf{59 (4.5\%)} \\
 && Others & 2 (0.5\%) & 5 (0.5\%) & \textbf{7 (0.5\%}) \\
 && Unknown & 156 (39.8\%) & 170 (18.6\%) & \textbf{326 (25.0\%)} \\
\midrule

&& {\textless{}=2} & {$35\ (8.9\%)$} & {-} & {-} \\
&& {$2-5$} & {$42\ (10.7\%)$} & {-} & {-} \\
\multicolumn{1}{l}{{\textbf{Disease duration in years}}} 
&& {$5-10$} & {$36\ (9.2\%)$} & {-} & {-} \\
{range: $1 - 27$} 
&& {$10-15$} & {$19\ (4.8\%)$} & {-} & {-} \\
{mean: $6.32$, sdt: $5.24$} 
&& {$15-20$} & {$8\ (2.0\%)$} & {-} & {-} \\
&& {\textgreater{}$20$} & {$3\ (0.8\%)$} & {-} & {-} \\
&& {Unknown} & {$249\ (63.5\%)$} & {-} & {-} \\

\midrule

\multirow{3}{*}{\textbf{Recording Environment}}
&& \textsf{Home Recorded} & 67 (17.1\%) & 585 (64.0\%) & \textbf{652 (49.9\%)} \\
&& \textsf{Clinical Setup} & 117 (29.8\%) & 235 (25.7\%) & \textbf{352 (27.0\%)} \\
&& \textsf{PD Care Facility} & 185 (47.2\%) & 85 (9.3\%) & \textbf{270 (20.7\%)} \\
\bottomrule
\end{tabular}%
}
\label{tab:demographics}
\end{table}

\begin{table}[!htbp]
\centering
\caption{\textbf{Dataset Demographics for Model Evaluation Splits.} This table presents the demographic breakdown of participants across different splits for model evaluation: (a) using a conventional random train-validation-test split, and (b) using a cross-environment split where \textsf{Clinical Setup} and \textsf{PD Care Facility} recording environment alternately serves as the test set, with the remaining environments combined for training and validation.}
\subcaptionbox{\label{tab:random-split}}{
\resizebox{\columnwidth}{!}{
\begin{tabular}{lrrrrrrr}
\toprule
\textbf{Cohort} & \multicolumn{1}{c}{\textbf{\# of Participants}} & \multicolumn{1}{c}{\textbf{\# of PD}} & \multicolumn{1}{c}{\textbf{\# of Female}} & \multicolumn{1}{c}{\textbf{\# of Non-White}} & \multicolumn{1}{c}{\textbf{\# of < 50}} & \multicolumn{1}{c}{\textbf{\# of \textsf{Clinical}}} & \multicolumn{1}{c}{\textbf{\# of \textsf{PD Care}}} \\ \midrule
Training Set & 916 (70.1\%) & 284 (31.0\%) & 481 (52.5\%) & 82 (9.0\%) & 71 (7.8\%) & 291 (31.8\%) & 214 (23.4\%) \\ 
Validation Set & 195 (14.9\%) & 54 (27.7\%) & 116 (59.5\%) & 19 (9.7\%) & 17 (8.7\%) & 25 (12.8\%) & 30 (15.4\%) \\ 
Test Set & 195 (14.9\%) & 54 (27.7\%) & 98 (50.3\%) & 18 (9.2\%) & 19 (9.7\%) & 36 (18.5\%) & 26 (13.3\%) \\ \bottomrule
\end{tabular}%
}}
\\[5mm] 
\subcaptionbox{\label{tab:generalizability-split}}{
\resizebox{\columnwidth}{!}{
\begin{tabular}{p{7cm}rrrrr}
\toprule
\textbf{Cohort} & \multicolumn{1}{c}{\textbf{\# of Participants}} & \multicolumn{1}{c}{\textbf{\# of PD}} & \multicolumn{1}{c}{\textbf{\# of Female}} & \multicolumn{1}{c}{\textbf{\# of Non-White}} & \multicolumn{1}{c}{\textbf{\# of < 50}} \\ \midrule
Training Set: \textsf{Home Recorded} and \textsf{PD Care Facility} & 771 (59.0\%) & 225 (29.2\%) & 432 (56.0\%) & 94 (12.2\%) & 77 (10.0\%) \\ 
Validation Set: \textsf{Home Recorded} and \textsf{PD Care Facility} & 183 (14.0\%) & 50 (27.3\%) & 105 (57.4\%) & 19 (10.4\%) & 21 (11.5\%) \\ 
Test Set: \textsf{Clinical Setup} & 352  (27.0\%) & 117 (33.2\%) & 158 (44.9\%) & 3 (0.9\%) & 9 (2.6\%) \\ \midrule
Training Set: \textsf{Home Recorded} and \textsf{Clinical Setup} & 837 (64.1\%) & 171 (20.4\%) & 448 (53.5\%) & 83 (9.9\%) & 67 (8.0\%) \\ 
Validation Set: \textsf{Home Recorded} and \textsf{Clinical Setup} & 199 (15.2\%) & 36 (18.1\%) & 106 (53.3\%) & 20 (10.1\%) & 15 (7.5\%) \\ 
Test Set: \textsf{PD Care Facility} & 270 (20.7\%) & 185 (68.5\%) & 141 (52.2\%) & 13 (4.8\%) & 25 (9.3\%) \\ \bottomrule
\end{tabular}%
}}
\end{table}

\subsection{Feature Extraction}
In this study, we aim to capture the nuanced speech dynamics indicative of PD using advanced deep learning embeddings from pangram utterance speech. 
We utilized three state-of-the-art semi-supervised speech models: Wav2Vec 2.0~\cite{baevski2020wav2vec}, WavLM~\cite{chen2022wavlm}, and ImageBind~\cite{girdhar2023imagebind} to extract intermediate vector representations from their last hidden layers, capturing sophisticated and informative features of the speech data.
Alongside, to assess the efficacy of these deep embedding features relative to traditional acoustic features, we extracted classical features using the methodology proposed by Rahman et al.~\cite{rahman2021detecting}. For comprehensive analysis, we compiled four distinct feature sets from our audio datasets: $39$-dimensional acoustic features, $768$-dimensional Wav2Vec 2.0 features, $1024$-dimensional WavLM features, and $1024$-dimensional ImageBind features. We then trained various deep learning models using these feature sets to distinguish between individuals with and without PD, exploring how different types of features contribute to the model's performance.

\subsection{Performance Evaluation on Standard Train-Validation-Test Split}
In our initial experiment, we combined all three data cohorts, segmenting the dataset into three random splits: $70\%$ for training, and $15\%$ each for validation and testing.  
The validation set was used to select the best-performing model for subsequent testing. 
Table~\ref{tab:random-split} provides the data split details and demographic distribution across the subsets, where we can observe that each subset has a fairly balanced representation of demographic subgroups. Importantly, the split was conducted based on participants' IDs, ensuring that all data samples from each participant were grouped into either the train, validation, or test sets. Furthermore, the stratified splitting method guaranteed fair representation of PD participants across all three sets.

\textbf{{Baseline Modeling}}.
{We developed multiple baseline models to benchmark our approach, incorporating both traditional and deep embedding features. First, we implemented a CNN-based model trained directly on raw speech data as an end-to-end PD classification pipeline. This model, however, underperformed with an AUROC of $59.18\%$ and accuracy of $63.71\%$, highlighting the challenges in capturing PD-specific speech characteristics directly from time-series audio data.
CNNs are architecturally constrained by their local receptive fields, which limits their ability to capture both global context and long-range dependencies in the input data~\cite{vaswani2017attention, wang2018non}, leading to suboptimal performance in this task.
}
{Additionally, we tested both support vector machine (SVM) and neural network classifiers across four distinct feature sets: classical acoustic features, WavLM, Wav2Vec2, and ImageBind embeddings.} 
{Among the SVM-based models using four different feature sets, the one trained on WavLM embeddings demonstrated the best performance, achieving an AUROC of $75.43\%$ and accuracy of $69.65\%$, surpassing the SVM model with classical acoustic features, which reached an AUROC of $74.82\%$ and accuracy of $67.94\%$.}

Our neural network-based classifier, particularly when trained with WavLM embeddings, significantly outperformed all other baseline models. It achieved an AUROC of $85.89\%$ and an accuracy of $81.01\%$, surpassing the closest competitor --- model trained with ImageBind features --- by $5\%$ in AUROC. Despite a comparatively lower sensitivity of $56.25\%$, the model excelled in other metrics, achieving a specificity of $90.63\%$, a Positive Predictive Value (PPV) of $81.01\%$, and the Negative Predictive Value (NPV) of $80.79\%$. 
Notably, all models trained with deep embedding features consistently surpassed those using traditional acoustic features, {and the neural network-based classifiers outperformed the SVM models in general. Please refer to the first segment of six rows in Table~\ref{tab:results} for a summary of the evaluation metrics across various baseline models.}

\textbf{{Fusion Modeling.}}
To enhance model performance by leveraging the complementary strengths of different feature sets, we developed fusion models using concatenation and projection-based approaches.  
By concatenating four different feature sets in all possible combinations --- resulting in 11 unique set --- we observed consistent improvements, especially with combinations involving WavLM features.
The best results were achieved when combining all three deep embeddings (Wav2Vec2, WavLM, and ImageBind), with an AUROC of $89.49\%$ and accuracy of $82.28\%$. Although specificity ($85.99\%$) and PPV (73.17\%) were slightly lower than the best baseline model, this approach significantly improved sensitivity (from $56.25\%$ to $75\%$) and NPV (from $80.79\%$ to $87.10\%$)

\begin{figure*}[t]
  \begin{minipage}{.25\textwidth}
    \caption{\textbf{Performance evaluation of PD classifiers from speech in a random split configuration.} (a) and (b) respectively demonstrate the AUROC curve and the confusion matrix of our best performing novel fusion model which projects WavLM features into the feature space of ImageBind features.}
    \label{fig:performance}
  \end{minipage}%
  \begin{minipage}{.05\textwidth}
  \phantom{text}
  \end{minipage}%
  \begin{minipage}{.7\textwidth}
    \begin{subfigure}[t]{.48\textwidth}
      \centering
      \includegraphics[width=\linewidth]{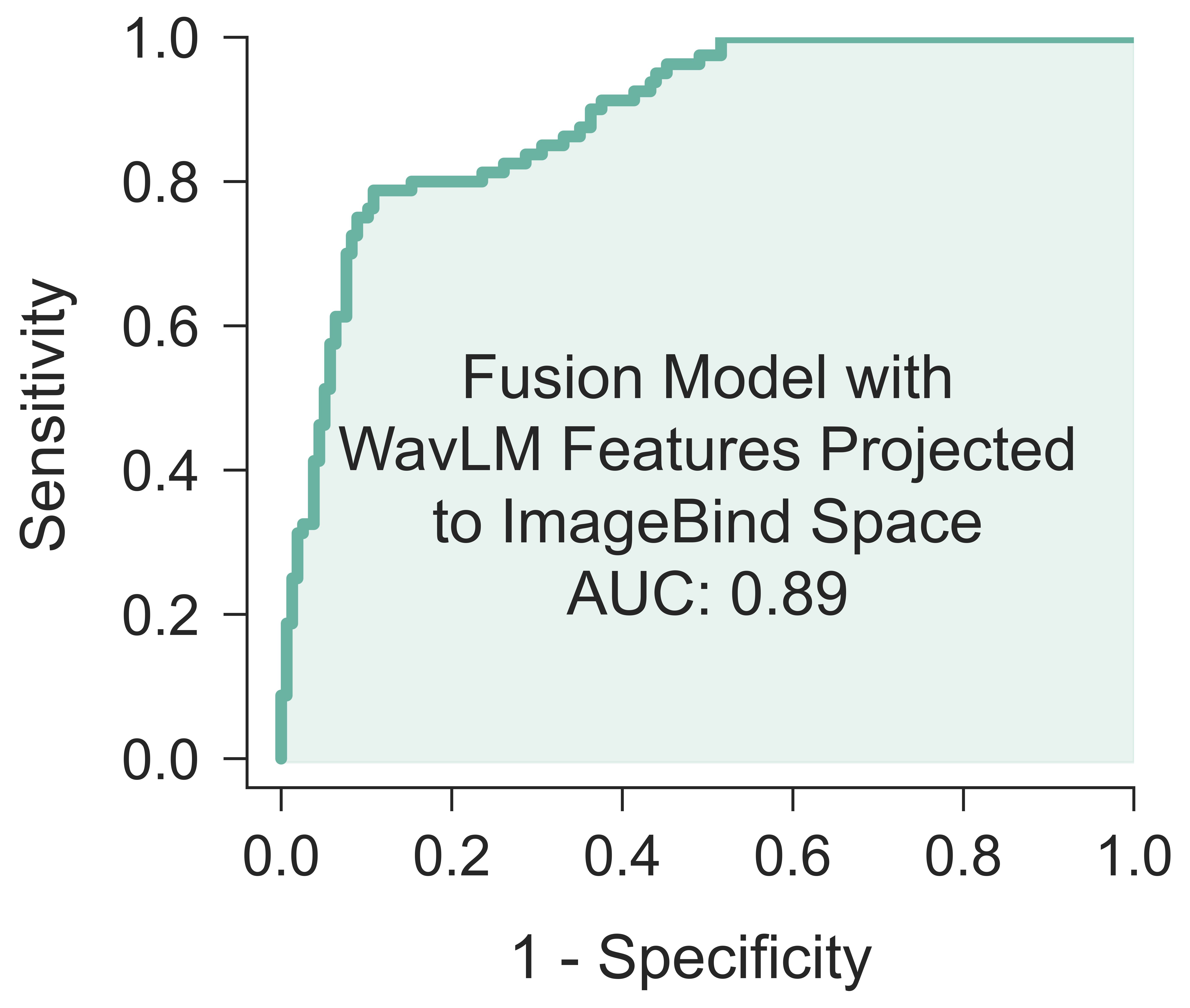}
      \caption{}
      \label{fig:roc_fusion_hybrid}
    \end{subfigure}%
    \begin{subfigure}[t]{.48\textwidth}
      \centering
      \includegraphics[width=\linewidth]{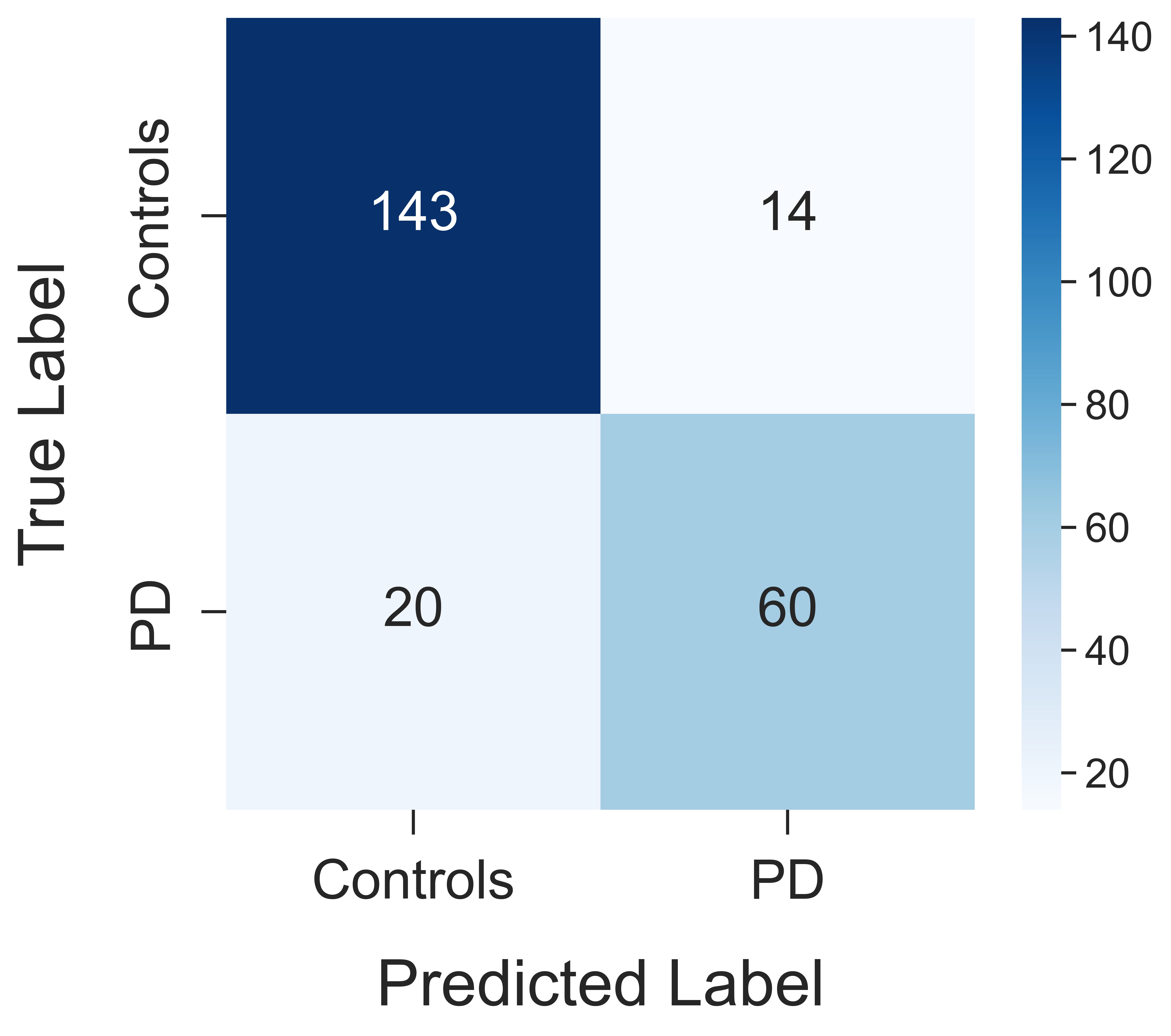}
      \caption{}
      \label{fig:cm_fusion_hybrid}
    \end{subfigure}
  \end{minipage}
\end{figure*}

In recent years, projection-based fusion has emerged as a powerful approach for enhancing representation learning in classifiers.
This method involves projecting features from one feature space into the space of another feature set, thus optimizing the use of complementary information while minimizing redundancy.
While our early fusion models using concatenation showed promising improvements, our projection-based fusion models further refined PD classification. 
Despite a slight decrease in AUROC ($88.94\%$ compared to the previous best of $89.49\%$), the fusion model projecting WavLM features into the feature space of ImageBind features achieved a significant increase in accuracy to $85.65\%$ from $82.28\%$. 
Additionally, it outperformed all other models (or achieved the similar best performance), in terms of other evaluation metrics with a sensitivity of $75.00\%$, specificity of $91.08\%$, PPV of $81.08\%$, and NPV of $87.73\%$. Figure~\ref{fig:performance} demonstrates the ROC curve and confusion matrix of this best-performing fusion model. 

{
Since our dataset was imbalanced, we evaluated our best fusion model on a resampled dataset using three different techniques: SMOTE~\cite{chawla2002smote}, Random Undersampling~\cite{mohammed2020machine}, and Random Oversampling~\cite{mohammed2020machine}. Among these, SMOTE yielded the best performance. However, even with the balanced dataset generated by synthetic samples, SMOTE failed to outperform the model developed without addressing data imbalance, achieving only an AUROC of $74.45\%$.
Furthermore, beyond evaluating the model on a randomly selected test set, we performed a $10$-fold cross-validation on our best-performing (on the randomly selected test set) model configuration. This configuration yielded similar performance, with an AUROC of $90.86\%$ and an accuracy of $85.37\%$. It also demonstrated $81.88\%$ sensitivity, $89.53\%$ specificity, $80.29\%$ PPV, and $88.63\%$ NPV, further reinforcing the model's robustness. The middle segment with six rows in Table~\ref{tab:results} summarizes the performance achieved by some of our top-performing fusion models.}

\begin{figure*}[t]
\vspace{10mm}
  \begin{subfigure}[t]{.45\textwidth}
    \centering
    \includegraphics[width=0.75\linewidth]{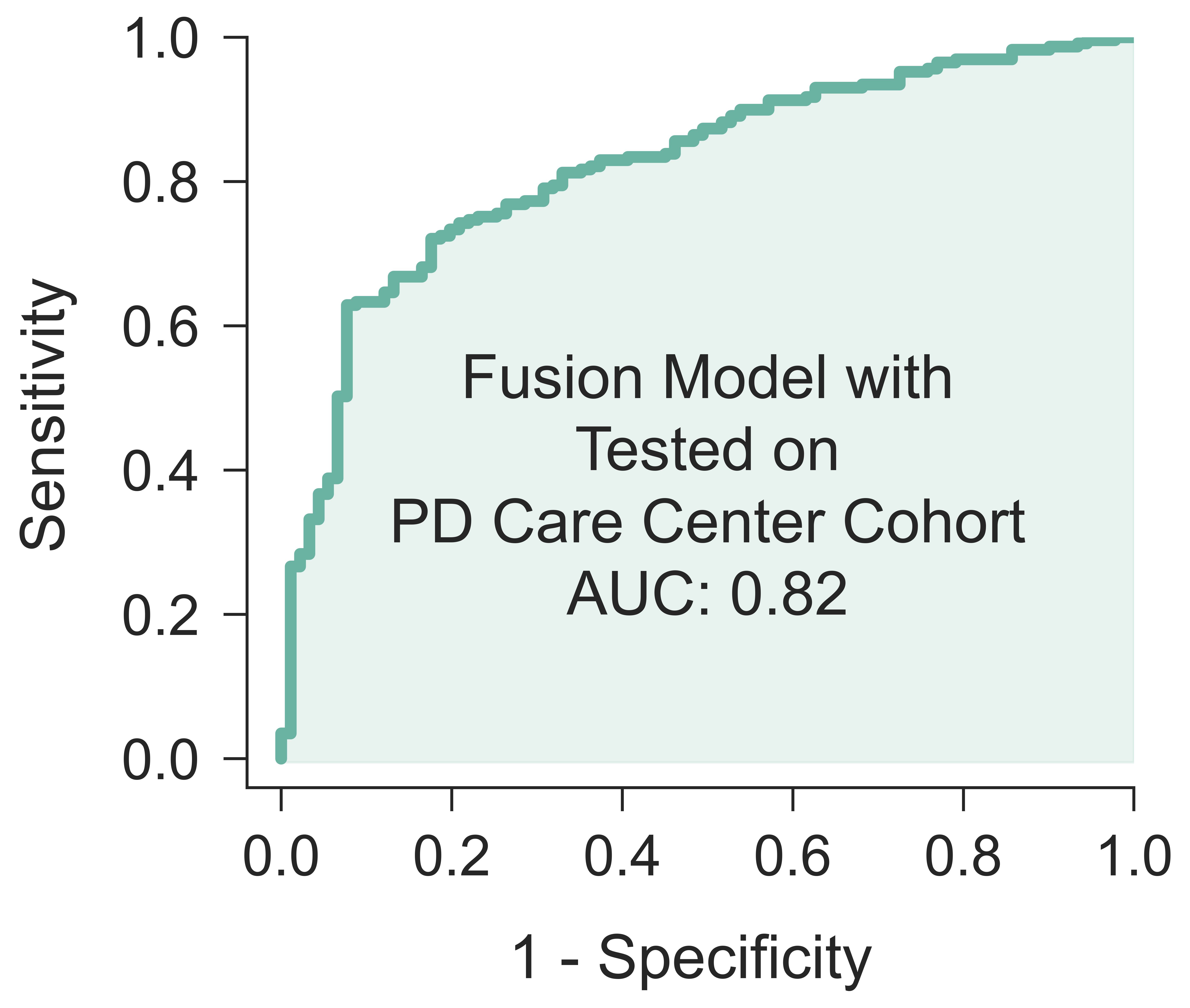}
    \caption{}
    \label{fig:roc_care}
  \end{subfigure}
  \hfill
  \begin{subfigure}[t]{.45\textwidth}
    \centering
    \includegraphics[width=0.75\linewidth]{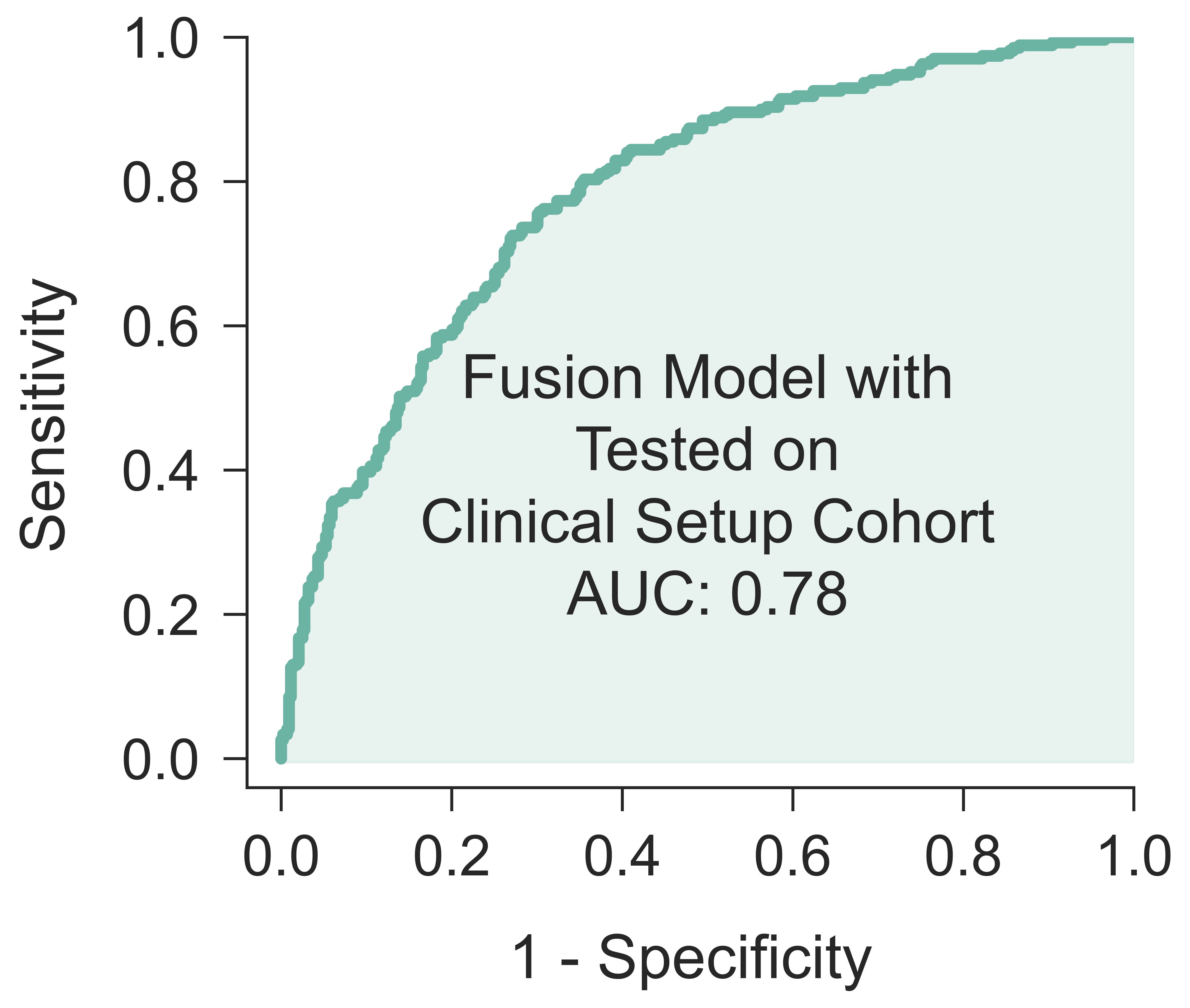}
    \caption{}
    \label{fig:roc_clinic}
  \end{subfigure}
  \bigskip
  \begin{subfigure}[t]{.45\textwidth}
    \centering
    \includegraphics[width=0.75\linewidth]{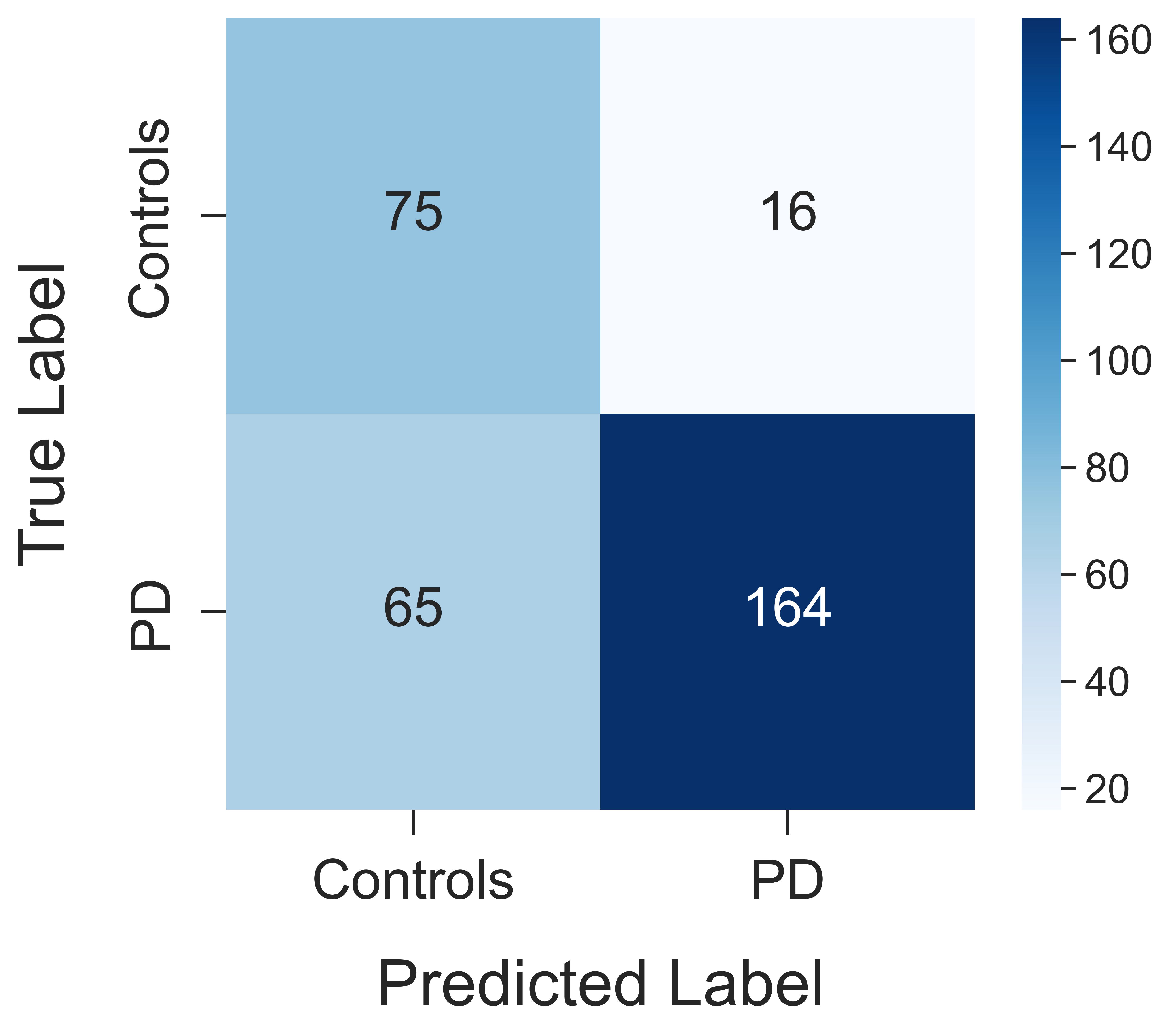}
    \caption{}
    \label{fig:cm_care}
  \end{subfigure}
  \hfill
  \begin{subfigure}[t]{.45\textwidth}
    \centering
    \includegraphics[width=0.75\linewidth]{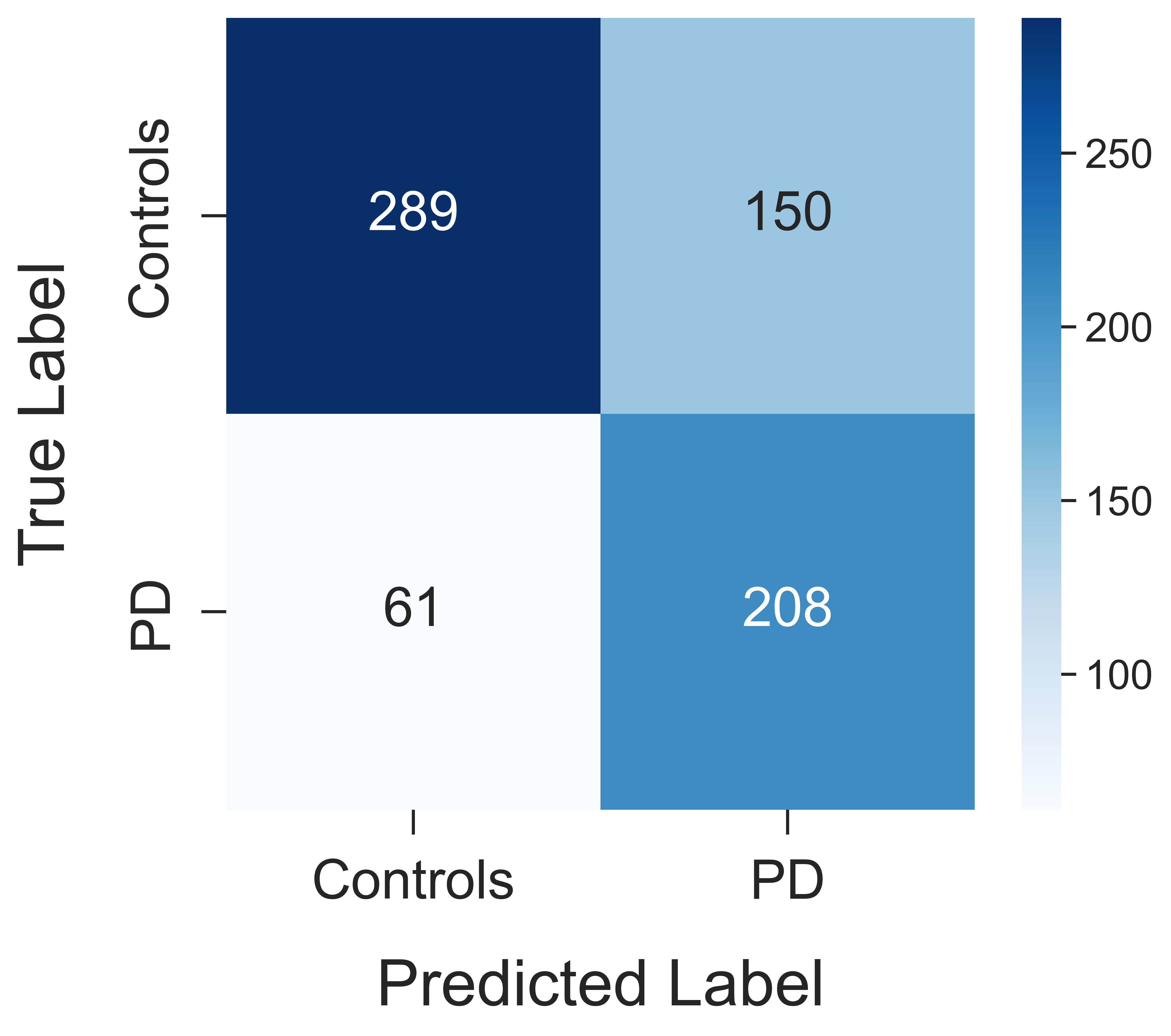}
    \caption{}
    \label{fig:cm_clinic}
  \end{subfigure}
  
    \caption{\textbf{Performance evaluation of PD classifiers from speech on external test sets.} (a) and (c) respectively demonstrate the AUROC curve and the confusion matrix of our best performing novel fusion model when tested on the dataset collected from \textsf{PD Care Facility}. In contrast, (b) and (d) give such visualizations when the model was tested on the participating cohort from \textsf{Clinical Setup}.}
  \label{fig:performance_external}
\end{figure*}

\subsection{Generalizability Test on External Datasets}

To evaluate the model's generalizability and its performance on datasets with probable distribution shift, we tested our best-performing fusion model using the datasets from \textsf{Clinical Setup} and \textsf{PD Care Facility} cohorts, while the training was conducted on the remaining cohorts, including the \textsf{Home Recorded} cohort. 
Due to a significant imbalance between PD and control participants in the \textsf{Home Recorded} cohort, with only $68$ PD participants compared to $585$ controls, it was not used in external testing. The demographic distribution of the training, validation, and test sets for this generalizability test is detailed in Table~\ref{tab:generalizability-split}. In the first configuration, using the \textsf{Home Recorded} and \textsf{Clinical Setup} cohorts for training and testing on the \textsf{PD Care Facility} cohort, the model achieved an AUROC of $82.12\%$ and accuracy of $74.69\%$. This represents a decline of $6.82\%$ in AUROC and $10.96\%$ in accuracy from the best random split performance. Additionally, the model showed a sensitivity of $71.61\%$, specificity of $82.42\%$, PPV of $91.11\%$, and a notably lower NPV of $53.58\%$.

In the second configuration, testing on the \textsf{Clinical Setup} cohort after training on the \textsf{Home Recorded} and \textsf{PD Care Facility} cohorts, the model recorded an AUROC of $78.43\%$ and accuracy of $70.19\%$, down by $10.51\%$ and $15.46\%$ respectively. Sensitivity was $77.32\%$, specificity was $65.84\%$, PPV was $58.10\%$, and NPV remained comparable at $82.58\%$. Figure~\ref{fig:performance_external} demonstrate the ROC curve and confusion matrix when our best fusion model was tested on data from the \textsf{Clinical Setup} and \textsf{PD Care Facility} cohorts, respectively. 
{The detailed evaluation metrics for these external datasets are shown in the last two rows of Table~\ref{tab:results}.}

\definecolor{green1}{HTML}{036D15}
\definecolor{green2}{HTML}{05A921}
\definecolor{green3}{HTML}{009900}
\definecolor{green4}{HTML}{05EC2B}

\begin{table*}[t]
\centering
\caption{{\textbf{Performance reporting for the ablation studies in a summarized version.} Note that a more detailed performance table is presented in the supplementary material.}}
\resizebox{1.0\linewidth}{!}{%
\tiny
\begin{tabular}{rrrrrrr}

\toprule

\multicolumn{1}{l}{{\textbf{Experimental Setup}}} & 
\multicolumn{1}{l}{{\textbf{AUROC}}} & 
\multicolumn{1}{l}{{\textbf{Accuracy}}} & 
\multicolumn{1}{l}{{\textbf{Sensitivity}}} & 
\multicolumn{1}{l}{{\textbf{Specificity}}} & 
\multicolumn{1}{l}{{\textbf{PPV}}} & 
\multicolumn{1}{l}{{\textbf{NPV}}} \\

\midrule

\multicolumn{1}{l}{{SVM w/ Acoustic Features}} & {74.82} & {67.94} & {64.19} & {69.93} & {53.06} & {78.67} \\
\multicolumn{1}{l}{{SVM w/ WavLM Embeddings}} & {75.43} & {69.65} & {64.19} & {72.50} & {55.31} & {79.28} \\
\multicolumn{1}{l}{{CNN w/ raw speech}} & {59.18} & {63.71} & {5.40} & {99.76} & {56.39} & {63.73} \\
\multicolumn{1}{l}{{Baseline w/ Classical Acoustic Feature}} & {72.79} & {69.95} & {61.54} & {75.20} & {60.75} & {75.81} \\
\multicolumn{1}{l}{{Baseline w/ WavLM Embeddings}} & {85.89} & {81.01} & {56.25} & {90.63} & {81.01} & {80.79} \\
\multicolumn{1}{l}{{Baseline w/ ImageBind Embeddings}} & {80.42} & {74.26} & {47.50} & {87.89} & {66.67} & {76.67} \\
\midrule

\multicolumn{1}{l}{{Concat WavLM, ImageBind, and Wav2Vec2}} & {89.49} & {82.28} & {75.00} & {85.99} & {73.17} & {87.10} \\
\multicolumn{1}{l}{{Concat all 4 feature sets}} & {87.91} & {78.82} & {61.54} & {89.60} & {78.69} & {78.87} \\
\multicolumn{1}{l}{{\textbf{Fusion w/ WavLM projected to ImageBind}}} & {88.94} & {\textbf{85.65}} & {\textbf{75.00}} & {\textbf{91.08}} & {\textbf{81.08}} & {\textbf{87.73}} \\
\multicolumn{1}{l}{{Fusion w/ all SSL embeds projected to 3rd dim}} & {\textbf{89.84}} & {81.01} & {70.00} & {86.62} & {72.73} & {85.00} \\
\multicolumn{1}{l}{{Fusion w/ WavLM projected to ImageBind (w/ SMOTE)}} & {74.45} & {73.42} & {58.66} & {86.67} & {64.47} & {76.58} \\
\multicolumn{1}{l}{{Fusion w/ WavLM projected to ImageBind (10--fold cv)}} & {90.86} & {85.37} & {81.88} & {89.53} & {80.29} & {88.63} \\
\midrule

\multicolumn{1}{l}{{\textsfsm{PD Care Facility} as Test Set}} & {82.12} & {74.69} & {71.62} & {82.42} & {91.11} & {53.57} \\
\multicolumn{1}{l}{{\textsfsm{Clinical Setup} as Test Set}} & {78.44} & {70.20} & {77.32} & {65.83} & {58.10} & {82.57} \\

\bottomrule
\end{tabular}
}
\label{tab:results}
\end{table*}

\subsection{Error Analysis}
To determine if the model is under performing for any particular demographic subgroup in the random split configuration, we performed rigorous error analysis in terms of four key demographic properties of the participants that were not included in the model's training and validation stages: sex, ethnicity, age, and recording environment.

\textbf{{Statistical Bias Analysis.}}
First, we performed statistical significance tests to evaluate whether the model's performance differed significantly across complementary demographic subgroups. {All statistical tests were conducted at an $\alpha = 0.05$ significance level.} 
{Given that we conducted six different hypothesis tests on the same evaluation set, we applied a Bonferroni correction~\cite{armstrong2014use} to control for multiple comparisons, adjusting the effective significance level to be $ \alpha^* = \frac{\alpha}{6} = 0.0083$.}
First, we divided the dataset into multiple subgroups based on sex (Male vs. Female), ethnicity (White vs. Non-White), age (Below 50 years vs. 50 years and Above), and recording environment (\textsf{Home Recorded} vs. \textsf{Clinical Setup} vs. \textsf{PD Care Facility}). Participants missing demographic data were excluded from respective analyses. Note that in the random split experimental setting, our test set had 237 audio samples. 
{
To carry out the statistical tests, we employed Fisher's Exact Test~\cite{upton1992fisher} to compare proportions between each pair of subgroups. This test was chosen due to its strength in handling categorical data and its suitability for our sample size, without the need to assume normality. Fisher's Exact Test provides an exact p-value, ensuring the robustness of our results even in cases where the central limit theorem's assumptions~\cite{kwak2017central} for other parametric tests, like the $Z$-test, may not fully hold.}

For the analysis based on sex, the model achieved an accuracy of $84.3\%$ for the male subgroup ($115$ samples) and $86.9\%$ for the female subgroup ($122$ samples). The p-value for this comparison was {$0.5847$}, indicating that the difference in model performance between male and female participant groups was not statistically significant. Ethnically, the model achieved an accuracy of $82.5\%$ for the White subgroup ($183$ samples) and $100.0\%$ for the Non-White subgroup ($18$ samples). The p-value for this comparison was {$0.0834$}, which was also not statistically significant. Age wise, the model achieved an accuracy of $100.0\%$ for participants below $50$ years old ($19$ samples) and $85.9\%$ for participants aged $50$ years and above ($199$ samples). With a p-value of {$0.1423$}, again this difference was not statistically significant. Since we had three distinct recording environments, we conducted statistical test for each of the pairs. The model achieved an accuracy of $91.0\%$ for the \textsf{Home Recorded} cohort ($133$ samples) and $72.2\%$ for the \textsf{Clinical Setup} cohort ($72$ samples). The p-value for this comparison was {$0.0010$}, indicating that the difference was statistically significant at the $95\%$ level. 
For the comparison between \textsf{Home Recorded} ($133$ samples, $91.0\%$ accuracy) and \textsf{PD Care Facility} cohorts ($32$ samples, $93.8\%$ accuracy), the p-value was {$0.9211$}, showing no significant difference.
Lastly, in the comparison between the \textsf{Clinical Setup} ($72$ samples, $72.2\%$ accuracy) and \textsf{PD Care Facility} cohorts ($32$ samples, $93.8\%$ accuracy) shows a p-value of {$0.0176$}, {which also came out to be statistically insignificant considering our corrected effective significance level.} 
The outcome of different statistical significance test are summarized in Table~\ref{tab:bias_analysis}.

Additionally, we examined the influence of disease duration on model performance by using Spearman's rank correlation among $110$ samples with known durations. The correlation coefficient ($\rho = 0.18$) suggested a weak positive correlation between model accuracy and disease duration, which was not statistically significant ($p = 0.0579$). This indicates that the duration of the disease does not significantly affect the model's effectiveness.

\renewcommand{\arraystretch}{1.2}
\begin{table}[t]
\centering
\caption{\textbf{Statistical analysis across demographic subgroups.}}
\resizebox{0.65\columnwidth}{!}{
\begin{tabular}{cccccc}
\toprule
\textbf{Demographic} &\multirow{2}{*}{\textbf{Groups}} &\multirow{2}{*}{\textbf{\# of samples}} &\textbf{Group} &\multirow{2}{*}{\textbf{p-value}} &\multirow{2}{*}{\textbf{Significant?}}\\
\textbf{Property} & & &\textbf{Accuracy} & &\\
\midrule
\multirow{2}{*}{Sex} & Male &115 &84.3 &\multirow{2}{*}{{0.5847}} &\multirow{2}{*}{No}\\
& Female &122 &86.9 & &\\
\midrule
\multirow2{*}{Ethinicity} & White &183 &82.5 &\multirow{2}{*}{{0.0834}} &\multirow{2}{*}{No}\\
& Non-White &18 &100.0 & &\\
\midrule
\multirow{2}{*}{Age} & Below 50 &19 &100.0 &\multirow{2}{*}{{0.1423}} &\multirow{2}{*}{No}\\
& 50 and Above &199 &85.9 & &\\
\midrule
Recording & \textsf{Home Recorded} &133 &91.0 &\multirow{2}{*}{{0.0010}} &\multirow{2}{*}{Yes}\\
Environment&\textsf{Clinical Setup} &72 &72.2 & &\\
\midrule
Recording & \textsf{Home Recorded} &133 &91.0 &\multirow{2}{*}{{0.9211}} &\multirow{2}{*}{No}\\
Environment&\textsf{PD Care Facility} &32 &93.8 & &\\
\midrule
Recording & \textsf{Clinical Setup} &72 &72.2 &\multirow{2}{*}{{0.0176}} &\multirow{2}{*}{{No}}\\
Environment&\textsf{PD Care Facility} &32 &93.8 & &\\
\midrule
\end{tabular}
}
\label{tab:bias_analysis}
\end{table}

\textbf{{Detailed cohort-based Error Analysis.}}
We conducted an extended error analysis using the Microsoft Error Analysis Framework~\cite{microsoft_error_analysis}  to assess the reliability and fairness of our fusion model across different demographic groups. This analysis, visualized through hierarchical decision tree maps and heat maps, identified specific cohorts with elevated error rates, indicating areas where the model's performance could be suboptimal.

\begin{figure*}[t]
  \begin{subfigure}[t]{0.85\textwidth}
    \centering
    \includegraphics[width=1\linewidth]{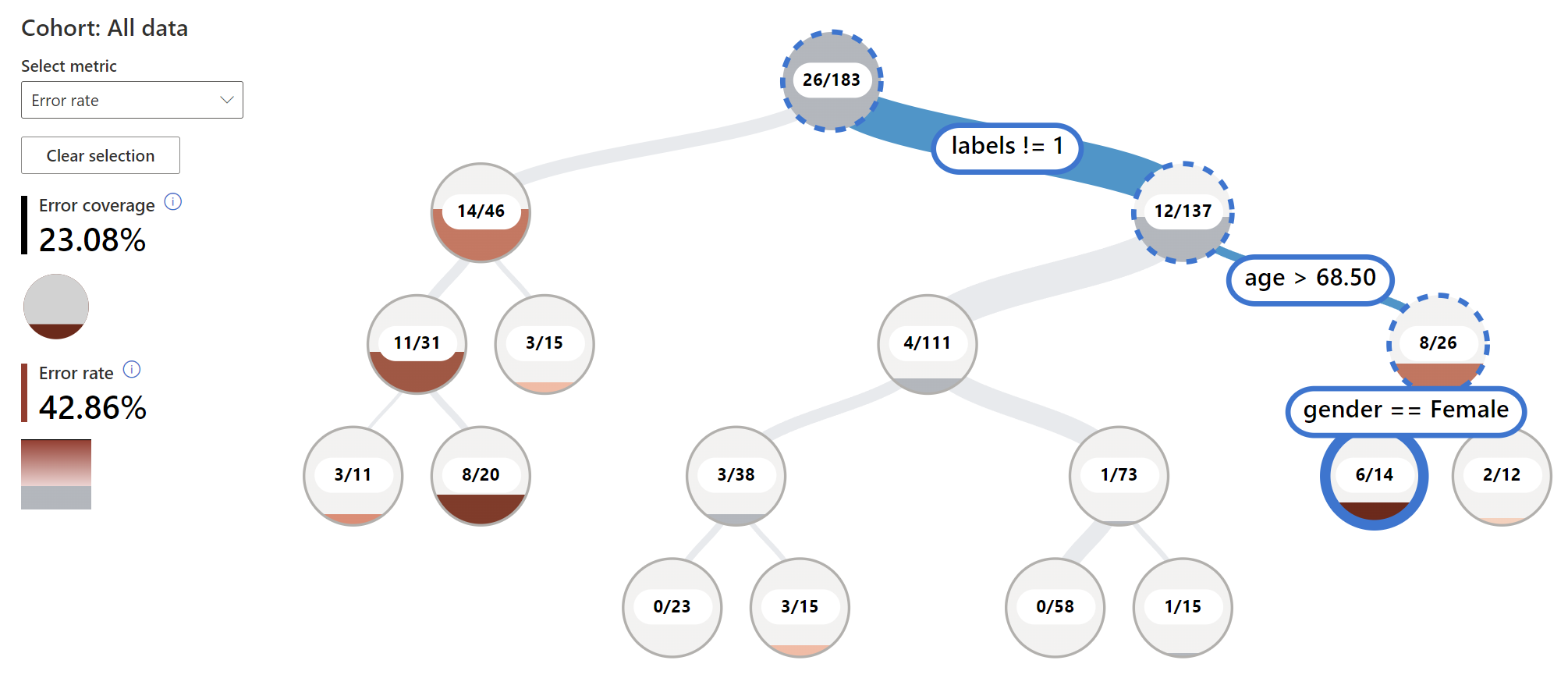}
    \caption{}
    \label{fig:cohort1}
  \end{subfigure}
  
  \begin{subfigure}[t]{0.85\textwidth}
    \centering
    \includegraphics[width=1\linewidth]{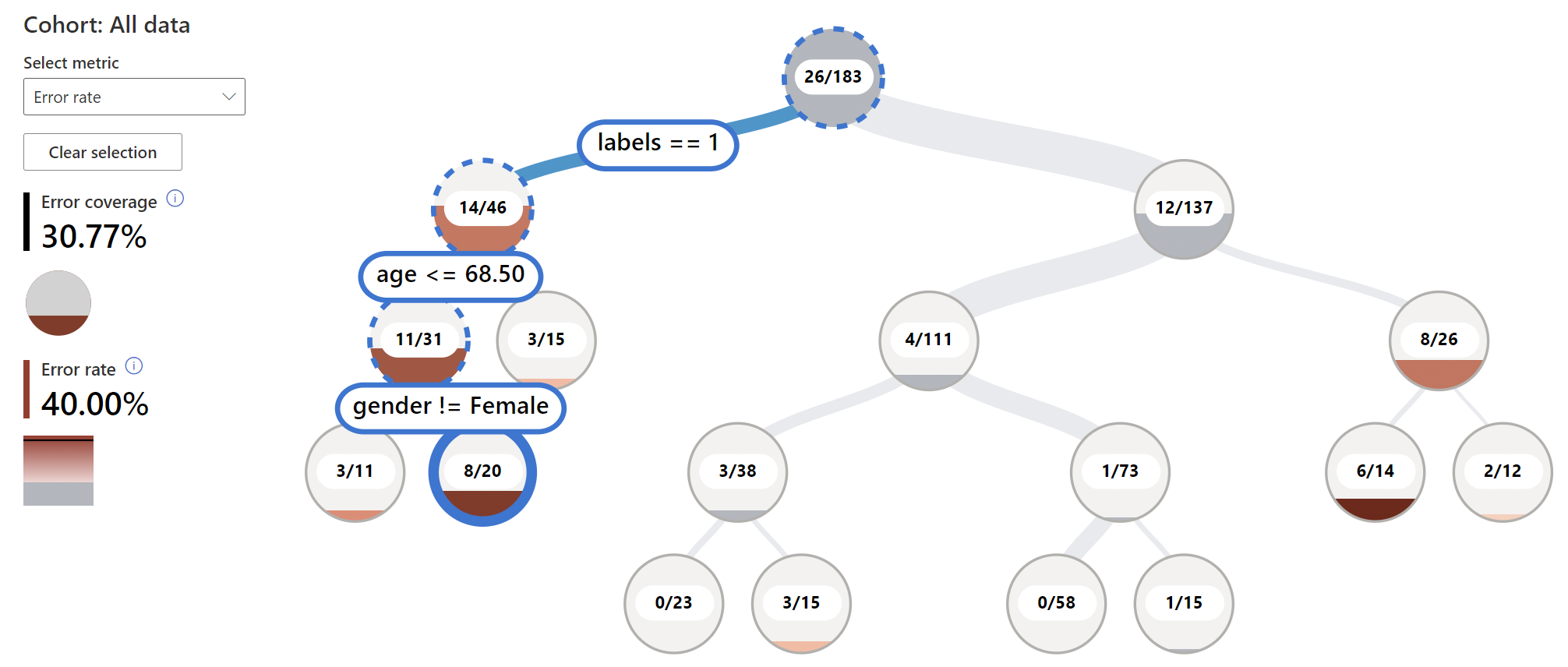}
    \caption{}
    \label{fig:cohort2}
  \end{subfigure}
    \caption{\textbf{Decision tree maps of error rates and error coverage among demographic cohorts.} (a) and (b) respectively demonstrate the notable nodes/cohorts with relatively high error rates and their error coverage percentage. The two numbers within each tree node represent the misclassified counts and the total counts of individuals in that specific cohort (i.e., $26/183$ indicates $26$ out of $183$ individuals were misclassified). Labels on the branch (i.e., age $\leq 68.50$) represent the decision boundary condition to split the child subtrees.}
  \label{fig:error_tree}
\end{figure*}

\textbf{Demographic Combinations:} From the decision tree visualization, we looked for cohort nodes with a stronger red color (representing a high error rate) branches with a higher fill line (representing high error coverage), and possible demographic combinations leading to errors. We identified a noteworthy control group of $26$ individuals (Figure~\ref{fig:cohort1}) aged above $68.5$ with a significant error rate and error coverage both of $30.77\%$. For its all-female subgroup ($14$ individuals), the model exhibited the lowest performance, 
with an error rate of $42.86\%$ and an error coverage of $23.08\%$. Besides this, we observed another PD cohort of $31$ individuals (Figure~\ref{fig:cohort2}) aged below $68.5$ with an error rate of $35.48\%$ and an error coverage of $42.31\%$, possessing a perfect PPV of $100\%$ but low sensitivity of $65\%$. For its all-male subgroup ($20$ individuals), the error rate increased to $40\%$ with an error coverage of $30.77\%$. These findings suggest that model adjustments towards these specific combinations of demographics could enhance accuracy and reduce misclassifications. 

\textbf{Age and Sex Disparities:} Our analysis highlighted significant age-related and sex-related disparities in error rates. From the heatmap visualization (detailed in the Supplementary Note 2 and Supplementary Figure 1, 2), we analyzed specific combinations of demographic features prone to higher error rates. Notably, white males aged $72.2-79.1$ ($21$ participants) and $51.5-58.4$ ($26$ participants) exhibited error rates of $33.33\%$ and $26.92\%$, respectively. PPV ($67\%$ and $57\%$) and sensitivity ($73\%$ and $50\%$) scores were low for these groups. Performance was suboptimal among females as well in these age ranges (error rates $60\%$ and $22.22\%$ respectively), indicating critical areas for model improvement.



\textbf{Ethnic and Sex Comparisons:} The model only erred for white individuals' classifications, presumably due to the small sample sizes of other ethnicity subgroups. White males ($72$ individuals) exhibited a higher error rate of $18.06\%$ compared to white females' ($93$ individuals) error rate of $13.98\%$, suggesting possible biases in the model.

\textbf{PD vs. Control Classifications:} We also examined the error differences between PD and control individuals across demographic subgroups. Notably, the model performed perfectly for a large control cohort ($44$ individuals) at an age interval of $58.4 - 65.3$ without any mistakes. The model erred more frequently for PD individuals aged between $51.5 - 58.4$ (error rate = $50\%$, error coverage = $15.38\%$). Control individuals were mostly misclassified for ages between $51.5 - 58.4$ (error rate = $15.79\%$, error coverage = $11.54\%$) and $77.2 - 79.1$ (error rate = $36.36\%$, error coverage = $15.38\%$).  PD white individuals were more often misclassified (error rate = $30.43\%$, error coverage = $53.85\%$) compared to control white (error rate = $10.08\%$, error coverage = $46.15\%$). Lastly, from the confusion matrix of sex versus PD labels, male PD individuals shared a slightly higher error rate of $32.14\%$ and error coverage of $34.62\%$ compared to female PD individuals' $27.78\%$ error rate and $19.32\%$ error coverage. Overall, for PD classifications, the model had an error rate of $30.43\%$ and contributed to $53.85\%$ of the total errors, whereas for control cases the error rate is only $8.76\%$. This indicates a need for enhanced PPV in detecting PD, as the PD group accounted for a majority of the misclassifications. The model appeared to be biased towards avoiding false positives. Therefore, improving model sensitivity and specificity for PD patients is crucial for reducing overall error rates and enhancing diagnostic accuracy.

In conclusion, these findings highlight the necessity for targeted interventions to address demographic-specific performance issues. This includes refining the model to reduce biases, enhancing training data diversity, and implementing demographic-specific adjustments to improve accuracy and fairness across all groups. The implications of these error patterns are critical for developing a more robust and equitable AI-driven healthcare solution.

\subsection{Ablation Studies}
Our ablation studies thoroughly evaluated the effectiveness of various feature sets for PD classification. 
We adopted different combinations of features sets for concatenation and also fused after projection to boost the metrics. For a detailed presentation of these results, please refer to the Supplementary Note 1 and Supplementary Table 1.

\section{Discussion}

In today's digital age, mobile devices have become pervasive across global populations, encompassing all age groups~\cite{taylor2023number}. These devices universally feature capabilities for audio recording, providing a practical platform for deploying our speech-based PD screening framework. By simply reciting a standard pangram, users can leverage their mobile devices to conduct preliminary screenings for PD. Our research further offers the potential to develop a mobile application utilizing semi-supervised speech models and fusion architecture and could continuously analyze natural speech during phone conversations --- with explicit user consent --- to detect early signs of PD and generate timely alerts. Such technological advancements significantly diminish the necessity for frequent clinical visits, offering a substantial benefit to individuals in areas where access to specialized neurological care is limited. This approach not only facilitates convenient at-home monitoring but also plays a crucial role in the early detection and managing the progression of PD, potentially altering the course of the disease by enabling earlier therapeutic intervention.

In 2017, PD placed a significant economic burden of \$52 billion in the United States~\cite{yang2020current}. Given the projected doubling of PD patients by 2030~\cite{dorsey2007projected}, the scenario will worsen significantly, surpassing \$79 billion even without accounting for inflation~\cite{yang2020current}. 
This growing financial strain underscores the crucial need for early diagnosis and regular monitoring, which have been shown to significantly improve patient outcomes and reduce the overall burden on healthcare~\cite{fujita2021effects}. 
Minimizing unnecessary clinical visits not only offers potential for substantial cost savings for both patients and healthcare providers but also means that each avoided visit --- which might otherwise be spent assessing healthy symptoms --- can save patients significant amounts on consultation fees and associated travel expenses.
The introduction of remote screening models like ours could therefore transform the management of PD, leading to a more efficient allocation of healthcare resources and financial savings.



Dashtipour et al.~\cite{dashtipour2018speech} highlighted that speech impairment affects up to $89\%$ of individuals with PD. In contrast to methods that require sustained phonation, the analysis of free-flow speech provides a more natural and comprehensive assessment of vocal impairments. This approach captures a wider spectrum of vocal characteristics and abnormalities, offering the potential for more accurate and earlier diagnosis than is possible with phonation-based models alone. 
Furthermore, the assessment of natural speech forms a core component of in-person evaluations as outlined by the MDS-Unified Parkinson's Disease Rating Scale (MDS-UPDRS)~\cite{goetz2008movement}, whereas sustained phonations are not explicitly monitored under these guidelines. While our study did not directly analyze continuous speech, the use of pangram utterance closely approximates natural speech patterns. The semi-supervised models employed in this research, which are trained on natural speech, are thus well-suited to capture the speech dynamics indicative of PD, even from the structured utterance of a pangram.

The projection-based fusion model has demonstrated superior performance compared to simple concatenation. 
The model projecting WavLM features into the ImageBind dimension achieved an AUROC of $88.94\%$ and accuracy of $85.65\%$, outperforming models using concatenated features, which despite a slightly higher AUROC of $89.49\%$, had lower accuracy at $82.28\%$. 
This result implies that projection-based fusion effectively aligns and synergizes different feature sets, overcoming issues like redundancy and scale disparity commonly seen with with concatenation. 
{Notably, this fusion method also surpasses models built on hand-crafted acoustic features, highlighting the limitations of traditional feature engineering in capturing the subtle, discriminative nuances of PD in speech. 
In contrast, embeddings derived from semi-supervised models like WavLM and ImageBind provably revealed complex PD-related speech patterns, suggesting that automated, embedding-based features significantly enhance the precision and depth of PD speech analysis -- a key contribution of our study.
}

Rizzo et al.~\cite{rizzo2016accuracy} reported that the accuracy of PD screening by non-expert clinicians stands at $73.8\%$, with a $95\%$ credible interval (CrI) of $67.8\% - 79.6\%$. In contrast, movement disorder specialists achieve a slightly higher accuracy of $79.6\%$, albeit with a wider CrI of $46\% - 95.1\%$. 
Encouragingly, our model demonstrates promising results, achieving an accuracy of $85.65\%$, positioning it favorably within or even above these established confidence intervals. {However, it is important to note that the populations used to evaluate our model and those in clinical studies are very different, so while this comparison is promising, it should be interpreted with caution.}
Our model's performance, coupled with its generalization capability across diverse recording environments, underscores its potential for global deployment among English-speaking populations. 
When tested on datasets from a PD care facility and a clinical setup, which were completely unseen during the training phase, our model demonstrated respectable AUROCs of $82.12\%$ (accuracy of $74.69\%$) and $78.44\%$ (accuracy of $70.20\%$), respectively. 
These results are comparable to, and in some cases surpass, those achieved by non-expert clinicians, highlighting the model's ability to deliver reliable PD screening.
{The observed drop in metrics during external validation can be attributed to variations in recording devices and environments,
indicating that machine learning models may be sensitive to environmental changes~\cite{bayram2022concept,ackerman2021machine}. 
However, this challenge also highlights a key strength of our approach compared to existing models, which are typically trained on data from a single recording device.
A distinctive characteristic of our home-recorded dataset is the diversity in recording devices, which mirrors real-world conditions better than controlled datasets. 
Nevertheless, we acknowledge that substantial differences in recording equipment can potentially degrade the model's performance. 
To mitigate the potential risk of overfitting to specific training environments and to ensure patients' safety, we recommend collecting a few initial data samples from the local environment at new deployment sites and retraining the model for optimal performance. 
In the future, we aim to further enhance the model's generalizability across diverse and unpredictable settings by continuously incorporating more data from unseen environments and retraining the model on increasingly heterogeneous data sources. While this study does not explore continual learning~\cite{hadsell2020embracing}, we plan to apply various continual learning algorithms in the future to effectively address the challenge of integrating new incoming data, ensuring the model remains adaptable and robust over time.
}



Our statistical significance tests show that the model demonstrates broad invariance to demographic diversity, with no significant bias detected across sex, ethnicity, or age subgroups.
{While we observed significant changes in performance across different recording environments, this finding further supports our objective of enhancing the model's generalizability by incorporating data from varied recording setups.
We also recognize that the analysis between the white and Non-White subgroups was limited in scope due to insufficient sample sizes of each of the Non-White ethnic groups. As a result, we were compelled to group all Non-White participants into a single category, which prevents a nuanced understanding of the model's performance across granular ethnic subgroups. 
This limitation affects the generalizability and reliability of the model for Non-White populations, leaving the findings somewhat inconclusive. 
Notably, this disparity in demographic representation, particularly within PD datasets, is a known challenge in the literature. For instance, nearly $90\%$ of the participants in the PPMI dataset~\cite{aleksovski2018disease} belong to the White ethnic group. Moving forward, we aim to address this limitation by collecting more data from underrepresented demographic subgroups and retraining the model incrementally to ensure optimal performance across diverse populations.}


Despite the model's overall robustness across demographic groups in terms of hypothesis testing, our detailed error analysis identified certain under-performing data cohorts with notable misclassification rates, indicating that we should be cautious interpreting the model's prediction results upon these cohorts.
While the model performed well for males under 51 and over 72 --- likely due to more distinct PD traits in these age brackets --- the transition age of 51 to 72 presents challenges. 
{The speech patterns in this middle age group are less distinct, possibly due to physiological changes that are harder for the model to interpret.}
{Subtle changes in speech in this age range may reflect both PD-related traits and other aging factors, making it harder for the model to distinguish the true source of variation.}
For females, the group over 72 also presents higher error rates, with 4 out of 6 errors being cases where the model predicted Non-PD participants to have PD. 
This misclassification may be due to significant physiological changes in vocal mechanisms, such as the thickening of the epithelium post-70~\cite{Chatterjee2011}, {which alters vocal characteristics and potentially confuses the model}. {The changes in speech patterns at higher ages for females might mimic anomalies associated with PD, causing the model to incorrectly identify them as PD cases.}
Studies support these observations, noting that males experience marked structural changes in their vocal mechanisms around age 60, while females undergo notable changes post-70. Males exhibit an increase in spectral energy skewness and nonlinear changes in fundamental frequency (F0) with age, whereas females show nonlinear changes in signal-to-noise ratio (SNR), further complicating model predictions in these age and gender groups \cite{Chatterjee2011, sammi2020, stathopoulos2011}. These variations in vocal mechanisms, along with sex-specific acoustic properties, likely contribute to the increased error rates observed in certain demographic groups in our study, warranting further exploration of these biological and neural factors in speech-based PD predictions.
Moreover, variations in misclassification linked to ethnic differences in speech patterns could be explored with a more ethnically diverse dataset, enhancing our understanding of the model's demographic discrepancies. 
{The overall sensitivity of our model is relatively lower at $75.0\%$, and it drops further to $70.0\%$ among the 46 PD participants with available demographic data. 
One of the contributing reasons for this higher false negative rate could be the imbalance in our dataset, where the number of PD participants is significantly lower than the Non-PD group
(another potential factor might be the model's sole reliance on speech, as discussed later in this section).
While data resampling techniques were considered to mitigate the data imbalance issue, the hyper-parameter tuning approach revealed that even effective approach, such as SMOTE was insufficient to improve the predictive performance, which coincides with previous findings that the higher dimensionality~\cite{chawla2002smote,blagus2013smote} of embeddings from models like WavLM and ImageBind can limit the efficacy of such techniques.
Although recruiting PD participants poses challenges due to logistical constraints, lower patient availability, and the complexities of remote data collection, in future, we plan to collaborate closely with clinics and research institutions, facilitating access to larger PD cohorts and ensuring a more balanced and representative dataset.
In the meantime, we advise cautious interpretation of the model's results to minimize any potential risks from misclassification, particularly with the higher false negative rates.}

One of the principal limitations we must acknowledge is the limited explainability of our model. 
While the use of vector embeddings from semi-supervised models like WavLM provides a powerful tool for feature extraction, the black-box nature of these models poses challenges in interpretability. 
{While tools like SHAP~\cite{lundberg2017unified} or LIME~\cite{ribeiro2016should} were considered for improving explainability, the abstract nature of embeddings---where dimensions lack inherent meaning---limits their utility.
At best, these tools could highlight which part of the embedding space contributes most to predictions, but this insight remains too abstract for clinical relevance.
We recognize the need for greater transparency of our model and plan to explore more suitable methods for model explainability in future work (potentially through novel approaches), as the field of explainable AI for embeddings is still evolving~\cite{gohel2021explainable,li2022explainability}.}
Additionally, our current reliance on English pangrams restricts the model's applicability to non-English speakers. However, the inherent adaptability of semi-supervised speech models offers a promising avenue for extending our approach to other languages.
{Recent research has shown significant progress in applying semi-supervised transfer learning to expand speech recognition models for low-resource languages.
Recent works have demonstrated the application of semi-supervised models for low-resource languages, such as methods for Italian~\cite{kim2021semi} and aphasic speech recognition in English and Spanish~\cite{torre2021improving}. Advances like 
using self-supervised representations for multilingual language diarization~\cite{frost2022fine} further illustrate the growing potential for handling multilingual and low-resource languages.
In future work, our group aims to explore fine-tuning or adapting existing models to generate speech embeddings for diverse languages. Given our model's relatively simple architecture and low-resource training requirements, once these models for other languages are available, we will be able to adapt and extend our approach to non-English speakers. This future direction not only highlights the potential for our model to accommodate speakers of different languages but also emphasizes the broader evolution of semi-supervised speech models beyond linguistic boundaries.}
As an preliminary analysis of our model's extendability, we took our best model trained only with pangram utterance data and tested it on a completely separate test dataset where the participants delivered a continuous, free-flowing speech for around one minute on their preferred topic (not any pangram). This test dataset involves 177 participants, and 39 of them are diagnosed with PD. Our model, even without seeing any such continuous speech, achieved a respectable AUROC of $77.4\%$ with an accuracy of $74.1\%$. 
This underscores the potential versatility of our approach in broader speech analysis contexts, not just for PD detection but possibly for other speech-related applications as well.
{In addition to the model's reliance on English language, a further challenge affecting its global applicability is the limited access to reliable technology, such as desktop or laptop computers with stable internet connections. Although addressing this constraint is beyond the scope of this study, we suggest that in regions where such resources are scarce at an individual level, deploying the tool at community-accessible locations equipped with necessary technology could help minimizing this barrier to some extent. While this setup may not fully achieve the level of accessibility we envision, it could nonetheless improve the tool's usability in underserved, remote areas.}

{The symptoms of PD vary widely among individuals, with some primarily exhibiting significant speech impairments, while others may experience motor symptoms like tremors without noticeable changes in their speech patterns.}
Consequently, our model, which primarily analyzes speech dynamics, may not be universally effective for all PD patients, especially those whose vocal symptoms are less pronounced early in the disease. 
{This could also contribute to the model's higher false negative rate.}
{To address this, we recognize the potential of integrating additional modalities into our framework.
Given the flexibility of being a web-based platform, incorporating video data collection for tasks like finger tapping (for motor assessment of bradykinesia~\cite{hallett1980physiological}) and facial expressions (for hypomimia~\cite{maycas2021hypomimia}) is feasible. 
Moving forward, we plan to incorporate these modalities into a unified model that considers speech, motor function, and facial expressions, offering a more comprehensive and reliable PD screening tool.
}
{This multimodal approach will allow the model to capture a broader spectrum of PD symptoms, improving the overall performance and utility across diverse patient groups.}
{In addition, validating the model's robustness across different stages of PD and analyzing the impact of MDS-UPDRS assessment scores for the speech task on model performance would have provided further insights.
However, collecting PD stage data and UPDRS scores in a home-based data collection procedure presents significant logistical challenges. Gathering this data requires substantial time and effort commitments from specialists, which we acknowledge as a limitation of the study. Moving forward, we plan to conduct a controlled data collection process where clinicians can be involved to provide UPDRS assessments with staging of the disease, allowing us to explore the relationship between speech-based model performance and disease severity more thoroughly.
}

{Integrating AI into healthcare brings significant ethical challenges, particularly regarding data privacy and the accuracy of assessments. To ensure data privacy, any deployed version of the tool should remove video and audio data immediately after feature extraction to mitigate unauthorized access risks, especially in large-scale clinical or at-home deployments. In a real-world scenario, another concern is the potential for inaccurate risk assessments, as false positives, observed in $8.92\%$ of cases, may cause unnecessary anxiety but could also prompt users to seek professional medical advice. To alleviate the distress, we recommend that the deployed system clearly inform users that the results are not a definitive diagnosis and not free from errors. Furthermore, integrating psychological care resources, such as support groups, would ensure users have access to emotional support if distressed by their results. Conversely, the $25\%$ false negative rate is more concerning as it could delay the necessary medical care. As we already discussed, this limitation may be due to dataset imbalance, as data from PD patient is harder to collect or the model's sole reliance on speech task, or perhaps both. Moving forward, we aim to address this by balancing the dataset and incorporating additional tasks, potentially enhancing model performance. Meanwhile, users must be informed of the tool's limitations on misclassification, emphasizing the importance of consulting healthcare providers regardless of the device's output.}

The implications of this study are broad, extending beyond PD diagnosis. The methodologies developed could be adapted for identifying other speech-related deficiencies, offering a blueprint for future research in neurological disorders. By integrating our proposed projection-based fusion architecture with semi-supervised speech embeddings, we anticipate similar methodologies could significantly improve the performance of models in various speech analysis applications. The success of this project highlights the transformative potential of AI in healthcare, particularly in enhancing diagnostic processes through advanced machine learning techniques and accessible digital platforms.

    

\section{Methods}

\subsection{Dataset Description}
\textbf{{Data Collection Framework.}} 
For collecting the speech dataset used in this study, we employed the web-based PARK framework developed by Langevin et al.~\cite{langevin2019park}, accessible at \url{https://parktest.net/}.
Participants worldwide can use this framework to record themselves while performing tasks inspired by the MDS-UPDRS guidelines designed to assess motor symptoms for evaluating PD.
One such task involved the articulation of a standard English pangram, ``quick brown fox.'' To ensure consistency and proper execution across participants, {the full pangram text was visible on the web interface for them to read aloud,} and an instructional video was provided before each task. Additionally, participants were required to complete questionnaires that captured demographic information such as age, sex, ethnicity, PD diagnosis year (for participants with PD), etc. From the video recordings, we extracted audio clips to compile our dataset. Due to some participants providing multiple video/audio samples at different times, we could amass a total of $1854$ samples for our study.
Additionally, to assess the extendability of our model to more natural speech scenarios from a pangram-only setting, we conducted a supplementary test. In this test, $177$ participants, including $39$ with PD, were instructed to speak freely for one minute on a topic of their choice. This additional dataset allows us to explore our model's performance in continuous free-flow speech settings, further underlining its potential applicability to broader speech analysis contexts. {Note that this study was approved by the Institutional Review Board (IRB) of the University of Rochester and the University of Rochester Medical Center. Informed consent was collected electronically due to the remote nature of the study, authorizing the use of participants' data and images.}\\

\textbf{{Data Collection Settings.}}
We collected the dataset from three distinct settings.
\begin{itemize}
    \item \textsf{Home Recorded}: We gathered a significant portion of our dataset from participants who recorded themselves staying at home using the PARK tool. We reached these participants by advertising our PARK tool on social media. We also emailed individuals who were willing to contribute to PD research. Despite being a global effort, we could only collect data from $67$ ($10\%$ of this cohort) PD participants. In this data collection setup, the labels of the participants (PD or control) were self-reported.
    \item \textsf{Clinical Setup}: In collaboration with the University of Rochester Medical Center (URMC) in New York, participants in a clinical study recorded themselves using the PARK tool. This setting ensured some supervision by clinical staff, particularly for those participants who required assistance during the recording. The participants of this cohort were clinically confirmed to be PD or control. Almost $30\%$ of the total PD participants are from this cohort.
    \item \textsf{PD Care Facility}: Our last data collection site was InMotion (details at \url{https://beinmotion.org/}), a PD care facility in Ohio, US. We could collect the video/audio data from their clients as clinically confirmed PD patients and their caregivers as self-reported controls. This environment provided a supportive setting for participants, typically involving assistance from their caregivers and/or the InMotion's staff during recording sessions. We collect the major portion of PD samples ($47\%$) from this setup. Note that the supplementary dataset of continuous free-flow speech involving 177 participants, including 39 with PD, was also gathered here.
\end{itemize}

\textbf{{Data Demographic Details.}}
In this study, we collected data across a broad spectrum of demographic groups, targeting a comprehensive analysis of PD across diverse populations. 
The dataset featured a balanced sex distribution with $53.2\%$ female and $46.6\%$ male participants but showed a predominance of white participants at $66\%$, with $25\%$ not disclosing their ethnicity. Participants ranged from $16$ to $93$ years old, with a majority aged $60-69$ years, which is significant as PD prevalence increases with age. Data collection occurred in varied settings to enhance external validity: $49.9\%$ at home, $27\%$ in clinical setups, and $20.7\%$ at a PD care facility, reflecting the accessible nature of our methodology. {Although information on PD stages was unavailable, disease duration data were collected from 143 participants. Among these, over half had been living with PD for less than five years, which is typically considered the early stage of disease progression.}

\textbf{{Data Cleaning and {Pangram Extraction}}} 
To enhance the quality and utility of our dataset, we implemented a comprehensive data cleaning and preparation process for videos captured under varied conditions.
Initially, we standardized the video format by converting all WEBM files to MP4, ensuring consistent fps and uniform metadata. We then isolated the audio from each video, converting it to a WAV format sampled at 16 kHz, which is optimal for detailed acoustic analysis.
Using the Whisper To precisely capture segments containing the target pangram, we used the Whisper model~\cite{radford2023robust}, which provided robust transcription and timestamping. 
{This process involved detecting occurrences of specific keywords within the pangram -- `quick', `brown, `fox', `dog,' and `forest.' Whisper generated start and end timestamps for each identified segment, and we defined each clip using the start timestamp of the first keyword and the end timestamp of the last keyword in the detected sequence.}
We also extended each audio segment by 0.5 seconds on either end to retain contextual audio cues, {such as breaths, subtle background noise, and speech transitions, which provide valuable context around speech boundaries and contribute to a more informative representation of natural speech patterns}. These extended segments were then saved in WAV format to create a comprehensive speech dataset for further analysis.


\subsection{Digital Speech Feature Extraction}
This study aims to objectively and quantitatively capture the nuanced speech dynamics that can be pivotal in differentiating PD characteristics. To achieve this, we extracted a series of classical acoustic features alongside advanced deep learning embeddings using state-of-the-art models such as Wav2Vec2~\cite{baevski2020wav2vec}, WavLM~\cite{chen2022wavlm}, and ImageBind~\cite{girdhar2023imagebind}. The remaining part of this subsection details the methodologies employed for each type of feature extraction.

\textbf{Classical Acoustic Features}
We extracted classical acoustic features proven to be crucial in the literature in characterizing speech disorders associated with PD~\cite{little2007exploiting,rahman2021detecting,canturk2016machine}. These features include Mel-frequency cepstral coefficients (MFCCs)~\cite{saldanha2023jitter,hawi2022automatic}, jitter~\cite{saldanha2023jitter,upadhya2017statistical}, shimmer~\cite{upadhya2017statistical,canturk2016machine}, and pitch-related metrics~\cite{rusz2011quantitative,liu2012vocal}:

\begin{itemize}
  \item \textbf{MFCCs:} MFCCs represent the short-term power spectrum of sound, based on a linear cosine transform of a log power spectrum on a nonlinear mel scale of frequency~\cite{rabiner1978digital}.
  \item \textbf{Jitter:} Measures frequency variations from cycle to cycle, offering insights into the stability of vocal fold vibrations~\cite{sataloff2017voice}.
  \item \textbf{Shimmer:} Quantifies amplitude variations, useful in assessing vocal fold closure inconsistencies~\cite{sataloff2017voice}.
  \item \textbf{Pitch:} This feature encapsulates the fundamental frequency, providing crucial information on the tonal aspects of the speech~\cite{titze1998principles}.
\end{itemize}
Tools such as Praat~\cite{boersma2001speak,styler2013using} and Parselmouth~\cite{jadoul2018introducing} were utilized to calculate these features from the digitized voice recordings of the participants.

\textbf{Deep Embedding Features.}
One of the significant contributions of our study is the use of deep embeddings of the speech audio extracted by three distinct pre-trained semi-supervised language (SSL) models: Wav2Vec 2.0 (W2V2), WavLM, and ImageBind. 
{Our selection of these models was motivated by their robustness in capturing nuanced audio features and their ability to handle noisy speech data in complex audio environments, which is especially relevant for detecting subtle speech variations associated with PD.}

\textbf{Wav2Vec 2.0 (W2V2)}~\cite{baevski2020wav2vec} was developed by Facebook AI, which utilizes a self-supervised learning framework to learn high-quality representations from raw audio waveforms. 
{By predicting masked segments within the audio context, Wav2Vec 2.0 captures essential speech dynamics and produces embeddings well-suited for identifying subtle acoustic patterns, which is particularly valuable for PD detection.}

\textbf{WavLM}~\cite{chen2022wavlm} from Microsoft was build upon Wav2Vec 2.0's architecture, which enhances the model's ability to handle diverse acoustic environments, including noisy backgrounds and overlapping speech, making it highly effective for applications in real-world scenarios. 
{This model's ability to adapt to unpredictable audio environments makes it effective for clinical or home-based PD detection, where environmental consistency cannot always be controlled, and its capacity to discern subtle audio patterns in noisy settings aligns well with the clinical requirements of PD speech analysis, offering refined acoustic detail extraction critical for our task.}


\textbf{ImageBind}~\cite{girdhar2023imagebind}, introduced by Meta AI, extends beyond conventional audio-only models by incorporating cross-modal learning between audio and visual data. By training on paired datasets of images and corresponding audio, the model learns to create embeddings that reflect not just the audio content but also its relation to visual elements, enhancing the ability to discern nuanced variations in speech possibly linked to neurological conditions like PD.

{Our decision to focus on these specific models over alternatives like UniSpeech~\cite{chen2021unispeech}, HuBERT~\cite{hsu2021hubert}, and XLS-R~\cite{babu2021xls} stems from their particular optimization goals. While these models are recognized for their strengths in multilingual and cross-lingual tasks, their emphasis on Automatic Speech Recognition (ASR) and speaker recognition does not fully align with our objective of extracting helpful acoustic high-resolution embeddings for PD detection where robustness to noise and subtlety in speech variation is more critical.} 

{To obtain deep embeddings, we fed each extracted pangram audio sample into our selected SSL models. These models process the raw audio input through multiple layers, progressively encoding complex representations of the speech data. We extracted the embeddings from the final hidden layer, capturing the most sophisticated, context-rich features of the audio.}
This focus on high-resolution, contextually robust embeddings provides a solid foundation for our analysis, empowering our approach to capture the nuanced acoustic anomalies linked with PD.



\subsection{{Feature} Pre-processing}
Before starting the training phase, several data pre-processing steps were employed to optimize the dataset for effective machine learning model training. 

{\textbf{Data Deduplication} was performed to ensure dataset integrity by removing duplicates based on data collected from the same participant on the same date, ensuring that each speech sample was unique and properly represented in the dataset. This process eliminates redundant observations, prevent any potential bias, maintains data quality, and avoids overrepresentation in model training.}

\textbf{Correlation Analysis} uses Pearson correlation coefficient~\cite{cohen2009pearson} to generate a correlation matrix among the feature set, and features exhibiting a correlation coefficient above a predefined threshold were identified and elimination. {This process mitigates multicollinearity, potentially enhancing model stability and interpretability while reducing dimensionality.} The threshold and the decision to drop correlated features were set as tunable parameters, allowing for flexibility and optimization based on the specific characteristics of our dataset. {Note that after after hyperparameter tuning, the best model eliminated correlated features when the coefficient exceeded a threshold of $0.85$.}



\textbf{Data Scaling} is essential to prevent features with larger scales from dominating model training, ensuring each feature contributes proportionally~\cite{jain2005data}. In this study, we explored two popular scaling methods --- Min-Max Scaling and Standard Scaling --- as they are well-suited for neural network-based architectures. {Min-Max Scaling, which maps features to a specified range (typically [0, 1]), is effective for features with varying ranges, promoting faster convergence by keeping all values within a controlled scale~\cite{patro2015normalization}. On the other hand, Standard Scaling, on the other hand, centers features around a mean of 0 with a standard deviation of 1, making it ideal for algorithms that benefit from normally distributed data or are sensitive to feature variance \cite{raschka2019python}.} We configured the choice of scaling method (including using it or not) as a hyperparameter, allowing optimization based on data characteristics and model requirements. {Min-Max Scaling is particularly useful when there are large disparities in feature ranges, while Standard Scaling is preferred when the data distribution approximates Gaussian \cite{han2022data, scholkopf2018learning}. Applying these scaling techniques contributes to stable training, especially for gradient-based algorithms, by reducing distributional disparities across training, validation, and test sets, thereby enhancing model robustness and generalizability \cite{lecun2002efficient}. After hyperparameter tuning, Standard Scaling was selected for the best model performance.}

\textbf{Data Resampling} was employed to address the dataset imbalance, as the proportion of PD samples was approximately half that of control samples across all cohorts. Imbalanced datasets can lead to biased models that overfit the majority class and underperform in predicting the minority class, which, in this case, was the PD class. To mitigate this, we applied {three resampling techniques:} the Synthetic Minority Over-sampling Technique (SMOTE)~\cite{chawla2002smote}, which generates synthetic samples from the minority class; {Random Undersampling~\cite{mohammed2020machine}, which reduces the control sample count to match the PD class; and Random Oversampling~\cite{mohammed2020machine}, which duplicates minority class samples to achieve balance.} We configured the choice of using one of these three resampling techniques alongside the option of not using any resampling as a tunable parameter, allowing a systematic evaluation of their impact on model performance. {However, after hyperparameter tuning, none of these techniques contributed to improved performance in the best model. As such, exploring and implementing more advanced data balancing techniques remains an area for future work, aiming to further mitigate the potential effects of data imbalance on model accuracy and robustness.}




\subsection{Baseline Modeling}

Following the setup of our data pre-processing and model development pipeline, we conducted initial baseline experiments. 
Our pre-processing pipeline, designed to be feature agnostic, enabled independent training of models on different feature sets (Classical, W2V2, WavLM, ImageBind). {To benchmark performance, we tested three primary model architectures: a Convolutional Neural Network (CNN) trained directly on raw speech data, Support Vector Machine (SVM) classifiers trained on extracted classical and deep embeddings, and deep learning models with fully connected layers also trained on these extracted feature sets. For the CNN model, we designed an end-to-end PD classification pipeline aiming to capture PD-specific characteristics directly from time-series data. In addition, we used SVM classifiers to explore linear and non-linear decision boundaries within each feature set.}

For our neural-network based baseline models, we employed two structures: a shallow fully connected classification layer (ShallowANN) with sigmoid activation and an Artificial Neural Network (ANN) with an additional hidden layer before the output layer. During the training of DL models, the choice between ShallowANN and ANN was kept as a hyperparameter to optimize performance based on the dataset characteristics and feature set.


\subsection{Fusion Modeling}

In our study, we explored several strategies to fuse multiple feature sets, enhancing the robustness and accuracy of the resulting models. We began with a simple vector concatenation approach, where distinct feature sets were merged into a single dataset. This basic concatenation strategy allowed us to establish a baseline for further fusion methods. Using the concatenated datasets, we trained both shallow and deep neural networks (ShallowANN and ANN) to assess the performance of combined features.

Moving beyond simple concatenation, we implemented a hybrid fusion approach, termed the \textit{``Projection-based Fusion Architecture''} in Figure~\ref{fig:overview}, which leverages projection and reconstruction techniques. Below, we describe the key components of this architecture.

{\textbf{Projection Layer} in our model architecture is essential for aligning multi-modal features within a shared latent space, enabling effective fusion of diverse feature types such as WavLM, Wav2Vec2, and ImageBind embeddings. 
This approach addresses the limitations inherent in simple concatenation, which often leads to feature mismatches, added noise, and increased dimensionality that can compromise model performance, particularly in high-dimensional, nuanced data like PD-related speech patterns~\cite{ngiam2011multimodal,baltrusaitis2019multimodal}.
By aligning features from different modalities, the projection layer enhances cross-modal interactions, making it possible for the model to leverage nuanced, multi-modal patterns crucial for detecting the subtle speech variations indicative of PD~\cite{veli2022transformer,wang2019alignment}.}

{Our choice to incorporate a projection layer is driven by several key considerations. 
First, by transforming features into a shared, lower-dimensional space, the projection layer reduces dimensionality and mitigates the risk of overfitting, which is especially important given the high-dimensional nature of speech features~\cite{hinton2006reducing}. 
Additionally, it enhances cross-modal alignment, allowing for deeper feature interactions that are essential for capturing PD-specific speech characteristics~\cite{lu2019vilbert}.
The layer also normalizes features across modalities, addressing scale and distribution differences that could hinder simple concatenation methods~\cite{zadeh2017tensor}.
This adaptive, learnable space enables dynamic feature fusion during training, promoting robust and generalizable representations -- a critical asset in PD detection where symptoms are subtle and varied~\cite{tsai2019multimodal}.}
{Our approach aligns with recent advancements in multimodal learning~\cite{lei2022univl,tan2019lxmert}, where projection layers are employed to unify diverse data types within a common space, as seen in models like the Multimodal Video Transformer (MMV)\cite{chen2022mm} and Multimodal Masked Autoencoder\cite{geng2022multimodal}. These examples underscore how projection techniques enhance joint representation learning, benefiting tasks such as action recognition and pretraining.}

{To optimize the projection layer, we conducted extensive hyperparameter tuning to find an ideal balance between information retention and model efficiency~\cite{bergstra2012random}. We experimented with a range of projection dimensions, from compact to more expansive, and further explored projecting one feature set into the space of another. Ultimately, the model achieved optimal performance by projecting WavLM features into the ImageBind feature space, which facilitated a richer, cross-modal alignment that was particularly proven to be effective in capturing the nuanced, multi-modal speech patterns relevant to PD detection~\cite{hlavnicka2017automated}.}

{\textbf{Fusion Layer} integrates the projected and aligned features into a cohesive, unified representation. First, by normalizing the projected features, this layer reduces distributional disparities across modalities. The normalized features are then combined, either by direct summation or by summing and further normalizing them, creating a unified cross-modal representation. This approach not only harmonizes diverse modality features but also enhances the model's ability to capture rich, multi-modal interactions, a critical aspect in tasks where nuanced multi-modal patterns are essential.}

{\textbf{Decision Layer} processes the unified representation to yield a final classification output. Two variations of the decision layer are employed as hyperparameter options: a deeper fully connected network and a simpler, shallow architecture. In the shallow configuration, the decision layer consists of a single fully connected layer, followed by a sigmoid activation, directly mapping the fused representation to a probability score. In contrast, the deeper variant introduces an intermediate layer, adding non-linearity and enabling the model to capture more intricate patterns in the fused representation before the final classification. The choice between these versions was tuned as a hyperparameter. The streamlined structure of both decision layers reduces model complexity and minimizes the risk of overfitting, particularly when working with limited data or high-dimensional features.}

{\textbf{Loss Function} for the projection-based fusion model is a multi-objective function with three components, each serving a crucial role in optimizing the model's performance and robustness:}

\begin{itemize}
    \item {\textbf{Prediction Loss}: This component uses Binary Cross-Entropy (BCE) to reduce the disparity between model predictions and ground truth, driving accuracy in classifying PD and non-PD cases \cite{de2005tutorial}.}

    \item {\textbf{Cosine Loss}: Calculated as the complement of cosine similarity between projected multi-modal features, this loss component guides the model to learn diverse, complementary representations~\cite{wang2018cosface}. It aims to capture unique aspects of each modality, potentially improving the model's ability to leverage multi-modal information.}

    \item {\textbf{Reconstruction Loss}: This loss ensures the fidelity of feature reconstruction from the projected space, using one chosen metric (hyperparameter) from Mean Squared Error (MSE), L1 norm, or Kullback-Leibler (KL) divergence~\cite{goodfellow2016deep}. The primary objective is to preserve essential input information while balancing dimensionality reduction and retention.}

\end{itemize}

{The combined weighted sum of these loss components enables fine-tuning of each objective's influence~\cite{kendall2018multi}, optimizing classification, feature representation, and information retention for a more robust and generalizable PD detection model.}

{
Our model architecture is intentionally designed to be relatively simple yet effective for PD detection, balancing performance with computational efficiency. 
Leveraging pre-trained models like WavLM, Wav2Vec2, and ImageBind for feature extraction allows our model to capture complex, high-dimensional representations without requiring additional deep layers, benefiting from transfer learning to enhance performance with limited PD-specific data~\cite{baevski2020wav2vec, chen2023imagebind, yosinski2014transferable}.
This approach reduces the risk of overfitting, which is common in complex architectures, especially with limited datasets in medical applications~\cite{hawkins2004problem}.
By maintaining an optimal data-to-parameter ratio, we improve generalizability and ensure stable performance across data splits~\cite{larochelle2009exploring, srivastava2014dropout}. 
Additionally, the simplicity of our model enables faster training and lower computational demands, aligning well with real-world clinical applications where efficiency and robustness are essential~\cite{sze2017efficient, zhang2018real}. 
Despite its streamlined design, our ANN strikes a balance between model complexity and generalizability, ensuring strong performance without unnecessary computational overhead. This is consistent with findings in prior studies, where even simple architectures --- such as linear layers --- have been shown to perform effectively for complex tasks like ECG-based AFib detection~\cite{diamant2022patient} and vision-language modeling~\cite{liu2023visual}, when paired with efficient pre-trained models to represent intricate data patterns. Inspired by efficient architectures like LLaVA, which successfully integrate multi-modal data, our model proves that high accuracy and stability can be achieved without excessive complexity~\cite{liu2023visual}.
}



\subsection{Evaluation Metrics}
To assess the performance of our deep learning models effectively, we used several key metrics that are crucial for clinical evaluation: Area Under the Receiver Operating Characteristic (AUROC) score, Accuracy, Sensitivity, Specificity, Positive Predictive Value (PPV), and Negative Predictive Value (NPV). 
We visualized the model's performance through AUROC curves and confusion matrices. The AUROC curve visually illustrates the model's discrimination ability between different conditions, and the confusion matrix details the counts of true positives, false positives, true negatives, and false negatives, essential for evaluating the model's practical efficacy.
The model achieving the highest AUROC score on the validation set was selected and then tested on the test set with these metrics to confirm its clinical relevance and effectiveness. 

\subsection{Statistical Bias Analysis}

To ensure equitable performance across diverse populations, we conducted a comprehensive bias analysis across demographic subgroups, specifically based on age, sex, ethnicity, and recording environment. Participants without demographic data were excluded from this analysis to maintain data integrity. {Given the multiple comparisons involved, we performed six distinct hypothesis tests and applied a Bonferroni correction~\cite{armstrong2014use} to control for Type I error, adjusting our effective significance level to $\alpha^* = \frac{0.05}{6} = 0.0083$.}





{For each comparison, we used Fisher's Exact Test to assess statistical significance. This test was chosen for its robustness with categorical data and its suitability for our sample size, as it does not require the normality assumption of parametric tests like the $Z$-test~\cite{kwak2017central}. Fisher's Exact Test provides an exact p-value, making it appropriate for detecting differences in performance proportions between subgroups without the reliance on large sample conditions. }

The bias analysis included subgroup comparisons by sex (Male vs. Female), ethnicity (White vs. Non-White), age (Below 50 vs. 50 and Above), and recording environment (\textsf{Home Recorded}, \textsf{Clinical Setup}, \textsf{PD Care Facility}). For recording environment, we conducted pairwise statistical tests between each setting to provide a detailed evaluation of how recording conditions may impact model performance. To further examine potential biases, we also analyzed the influence of disease duration on model performance by calculating Spearman's rank correlation among participants with known disease durations. This non-parametric test was selected to measure the association between model accuracy and disease duration without assuming a linear relationship.

\subsection{Cohort-based Error Analysis}

Building on the statistical bias analysis which assessed model fairness across demographics, our cohort-based error analysis delved deeper into individual and group-level performance nuances. 
This fusion model utilizes deep learning architecture and fusion features, which generally offer limited intuitive interpretability and explanation of the model's decisions, as well as challenges in identifying when and why the model errs.
Post-assessment using AUROC and other standard metrics, we tracked both predicted and true PD labels to explore the model's operational dynamics under varying demographic influences. We employed Microsoft's Error Analysis SDK~\cite{microsoft_error_analysis} to visually dissect error patterns and concentrations, particularly focusing on cohorts with heightened error occurrences. The error analysis process involved:
\begin{itemize}
    \item Cleaning the data by removing entries with missing demographic information (age, sex, ethnicity) and consolidating the test set to include necessary true and predicted labels and demographic features for each participant.
    \item Employing a decision-tree-like hierarchical structure to systematically identify error instances and understand prevalent error patterns by demographic features.
    \item Segregating the data into cohorts based on combinations of demographic characteristics to explore correlations with high prediction errors. Error rates, coverage, PPV, and sensitivity were visualized through heat maps for combinations -- \{sex $\times$ ethnicity\}, \{sex $\times$ age\}, \{age $\times$ ethnicity\}, \{sex $\times$ true PD labels\}, \{age $\times$ true PD labels\}, \{ethnicity $\times$ true PD labels\}.
\end{itemize}

This structured approach involving decision tree maps and heat maps allowed us to pinpoint under-performing demographic subgroups and potential triggers, enhancing our understanding of the model's performance across diverse populations.

\subsection{Hyperparameter Tuning}
Hyperparameter tuning~\cite{victoria2021automatic,feurer2019hyperparameter} is crucial for optimizing machine learning models to enhance their predictive capabilities. In our study, we rigorously tuned both the Baseline Modeling and Fusion approaches with the goal of maximizing the Area Under the Receiver Operating Characteristic (AUROC) for the validation set, a key metric for model selection. We employed Weights \& Biases (WandB), an advanced tool that supports systematic exploration of parameter space using a Bayesian optimization approach. This strategy enabled us to iteratively test numerous hyperparameter combinations to identify the configurations that delivered optimal performance. {Details of these hyperparameter settings with their search ranges and the chosen parameter by the best performing model are provided in the Supplementary Table 2 and 3, respectively  (Supplementary Note 3)}.

\section*{Code and Data Availability}
In accordance with the Health Insurance Portability and Accountability Act (HIPAA), we are unable to distribute the original audio recordings as they might reveal identifiable information about the participants. Nevertheless, upon acceptance of this manuscript, we will make publicly available the code for the feature extraction pipeline, the de-identified extracted features, and the model training scripts, thereby supporting transparency and reproducibility of our research.

\section*{Acknowledgement}
We express our profound gratitude to Meghan Pawlik for her substantial contributions to the maintenance of the clinical studies and the data collection. Cathe Schwartz and Karen Jaffe from the designated PD Care Facility were instrumental in gathering data. Amir Zadeh provided pivotal guidance throughout the project, helping to steer its direction and focus. Additionally, we are thankful to Sangwu Lee for his meticulous feedback on the manuscript and his invaluable assistance in debugging the code.

\section*{Competing Interests}
All the authors declare that there are no competing interests.

\section*{Author Contributions}
TA, AA, and MSI conceptualized the study and designed the experimental framework. AA and MSI were responsible for developing the feature extraction pipelines, while also collaborating on the development of the baseline machine learning models using individual feature sets. AA crafted the fusion models using concatenation techniques, whereas TA and EH$'$ (Ekram Hossain) focused on advancing the projection-based fusion architecture. Together, AA, TA, and EH$'$ engaged in model evaluation, results analysis, visualization, and manuscript preparation. RL and SP significantly contributed to error visualization and statistical analysis. All authors critically reviewed the manuscript, providing substantial revisions and improvements as needed. EH (Ehsan Hoque), as the principal investigator, along with MSI, supervised the entire project, providing strategic direction and refining the manuscript's narrative.

\bibliography{sample}

\newpage
\section*{Supplementary Materials}

\section*{Supplementary Note 1 -- Ablation Study}
We conducted extensive ablation studies to assess the discriminatory power of different feature sets in distinguishing participants with PD from the control group. 
We started with developing four deep learning baseline models, each trained and evaluated on one of four distinct feature sets.
We further investigated all possible combinations of these feature sets, constructing 11 models using concatenated heterogeneous features. 
Particularly, WavLM and ImageBind features showed promise, both individually and in combination, prompting us to explore performance enhancements through feature space projection.  
Notably, classical acoustic features were also integrated in these models as they might provide complementary information when projected and fused with deep embeddings.
Additionally, we projected both WavLM and ImageBind features (and Wav2Vec 2.0 in one instance) into a third, co-located latent feature space, subsequently fusing them to create a 512--dimensional feature set for PD classification.
Our findings revealed that the model projecting WavLM features into the ImageBind feature dimension outperformed all other configurations. Although the model that projects all three deep embeddings into a 512--dimensional latent space demonstrated the highest (slightly higher than our chosen best), we could not select it as the overall best performing model for being relatively lower in all other metrics. 
Consequently, our chosen optimal model was solely employed in subsequent external validations as it was tested on two unseen test datasets. Supplementary Table~\ref{tab:results} presents the evaluation metrics for all experimental setups implemented in our study.

\newpage

\definecolor{green1}{HTML}{036D15}
\definecolor{green2}{HTML}{05A921}
\definecolor{green3}{HTML}{009900}
\definecolor{green4}{HTML}{05EC2B}

\renewcommand{\arraystretch}{1.2}

\begin{table*}[!htbp]
\centering
\caption{\textbf{Performance reporting for the ablation studies.} Note that in the Fusion with Projection section, $A \rightarrow B$ means $A$ features set is projected to the dimension of $B$ features set. On the other hand the notation $A \rightarrow n, B \rightarrow n$ means both features sets A and B are projected to a their $n$ dimensional feature space.}
\resizebox{1.0\linewidth}{!}{%
\tiny
\begin{tabular}{rrrrrrr}
\arrayrulecolor{green2}\toprule
\multicolumn{1}{l}{\textcolor{green1}{\textbf{Experimental Setup}}} & \multicolumn{1}{l}{\textcolor{green1}{\textbf{AUROC}}} & \multicolumn{1}{l}{\textcolor{green1}{\textbf{Accuracy}}} & \multicolumn{1}{l}{\textcolor{green1}{\textbf{Sensitivity}}} & \multicolumn{1}{l}{\textcolor{green1}{\textbf{Specificity}}} & \multicolumn{1}{l}{\textcolor{green1}{\textbf{PPV}}} & \multicolumn{1}{l}{\textcolor{green1}{\textbf{NPV}}} \\
\arrayrulecolor{green2}\midrule
\arrayrulecolor{green4}\midrule
\multicolumn{7}{l}{\textcolor{green3}{\textbf{Deep Learning Baselines (Same NN based Classification Architecture on Different Feature Set)}}} \\
\arrayrulecolor{green4}\midrule
\multicolumn{1}{l}{Acoustic Feature} &  72.79& 69.95 & 61.54 & 75.20 & 60.75 & 75.81 \\
\multicolumn{1}{l}{Wav2Vec2 Embeddings} & 75.52 & 71.72 & 51.25 & 82.13 & 59.49 & 76.78 \\
\multicolumn{1}{l}{{WavLM Embeddings}} & {85.89} & {81.01} & {56.25} & {90.63} & {81.01} & {80.79} \\
\multicolumn{1}{l}{ImageBind Embeddings}& 80.42 & 74.26 & 47.50 & 87.89 & 66.67 & 76.67 \\
\arrayrulecolor{green4}\midrule
\multicolumn{7}{l}{\textcolor{green3}{\textbf{Other Baselines}}} \\
\arrayrulecolor{green4}\midrule
\multicolumn{1}{l}{{SVM w/ Acoustic Features}} & {74.82} & {67.94} & {64.19} & {69.93} & {53.06} & {78.67} \\
\multicolumn{1}{l}{{SVM w/ WavLM Embeddings}} &  {75.43}& {69.65} & {64.19} & {72.50} & {55.31} & {79.28} \\
\multicolumn{1}{l}{{CNN w/ Raw Speech}} & {61.50} & {64.02} & {5.36} & {97.60} & {56.10} & {64.31} \\
\multicolumn{1}{l}{{WavLM Fine-Tuning}} & {52.09} & {40.80} & {92.30} & {11.89} & {37.03} & {73.33} \\
\arrayrulecolor{green4}\midrule
\multicolumn{7}{l}{\textcolor{green3}{\textbf{Fusion Models (Concatenation of Fused Feature Sets)}}} \\
\arrayrulecolor{green4}\midrule
\multicolumn{1}{l}{Acoustic + Wav2Vec2} & 76.39 & 72.41 & 62.82 & 78.40 & 64.47 & 77.17 \\
\multicolumn{1}{l}{Acoustic + WavLM} & 84.56 & 75.37 & 57.69 & 86.40 & 72.58 & 76.60 \\
\multicolumn{1}{l}{Acoustic + Imagebind} & 79.28 & 72.91 & 66.67 & 76.80 & 64.20 & 78.69 \\
\multicolumn{1}{l}{Wav2Vec2 + WavLM} & 84.30 & 75.11 & 46.25 & 89.81 & 69.81 & 76.63 \\
\multicolumn{1}{l}{Wav2Vec2 + ImageBind} & 81.04 & 72.57 & 56.25 & 80.89 & 60.00 & 78.40 \\
\multicolumn{1}{l}{{WavLM + ImageBind}} & 89.39 & 81.01 & 60.00 & 91.02 & 78.69 & 81.82 \\
\multicolumn{1}{l}{Acoustic + WavLM + ImageBind} & 88.10 & 80.30 & 64.10 & 90.40 & 80.65 & 80.14\\
\multicolumn{1}{l}{Acoustic + WavLM + Wav2Vec2} & 80.03 & 71.92 & 51.28 & 84.80 & 67.80 & 73.61\\
\multicolumn{1}{l}{Acoustic + ImageBind + Wav2Vec2} & 79.73 & 70.44 & 52.56 & 81.60 & 64.06 & 73.38\\
\multicolumn{1}{l}{WavLM + ImageBind + Wav2Vec2} & {89.49} & 82.28 & \textbf{75.00} & 85.99 & 73.17 & 87.10\\

\multicolumn{1}{l}{All features} & 87.91 & 78.82 & 61.54 & 89.60 & 78.69 & 78.87\\

\arrayrulecolor{green4}\midrule
\multicolumn{7}{l}{\textcolor{green3}{\textbf{Fusion Models (with Projection)}}} \\
\arrayrulecolor{green4}\midrule
\multicolumn{1}{l}{WavLM $\rightarrow$ ImageBind} & 88.94 & \textbf{85.65} & \textbf{75.00} & \textbf{91.08} & \textbf{81.08} & \textbf{87.73} \\
\multicolumn{1}{l}{ImageBind $\rightarrow$ WavLM} & 88.20 & 81.43 & 71.25 & 86.62 & 73.08 & 85.53 \\
\multicolumn{1}{l}{WavLM $\rightarrow$ Acoustic} & 86.93 & 77.83 & 69.23 & 83.20 & 72.00 & 81.25 \\
\multicolumn{1}{l}{Acoustic $\rightarrow$ WavLM} & 85.60 & 75.86 & 67.95 & 80.80 & 68.83 & 80.16\\
\multicolumn{1}{l}{WavLM $\rightarrow$ 512, ImageBind $\rightarrow$ 512} & 88.63 & 78.90 & 67.50 & 84.71 & 69.23 & 83.65 \\
\multicolumn{1}{l}{WavLM $\rightarrow$ 512, ImageBind $\rightarrow$ 512, Wav2Vec2 $\rightarrow$ 512} & \textbf{89.84} & 81.01 & 70.00 & 86.62 & 72.73 & 85.00 \\
\multicolumn{1}{l}{{WavLM $\rightarrow$ ImageBind (10--fold cv)}} & {90.86} & {85.37} & {81.88} & {89.53} & {80.29} & {88.63} \\
\multicolumn{1}{l}{{WavLM $\rightarrow$ ImageBind (w/ SMOTE)}} & {74.45} & {73.42} & {58.66} & {86.67} & {64.47} & {76.58} \\
\multicolumn{1}{l}{{WavLM $\rightarrow$ ImageBind (w/ Random Oversamp)}} & {73.95} & {71.90} & {61.80} & {79.90} & {66.50} & {74.90} \\
\multicolumn{1}{l}{{WavLM $\rightarrow$ ImageBind (w/ Random Undersamp)}} & {71.50} & {70.60} & {57.20} & {84.20} & {59.80} & {76.50} \\
\arrayrulecolor{green4}\midrule
\multicolumn{7}{l}{\textcolor{green3}{\textbf{Generalizability Tests (with best Fusion Model)}}} \\
\arrayrulecolor{green4}\midrule
\multicolumn{1}{l}{\textsfsm{PD Care Facility} as Test Set} & 82.12 & 74.69 & 71.62 & 82.42 & 91.11 & 53.57 \\
\multicolumn{1}{l}{\textsfsm{Clinical Setup} as Test Set} & 78.44 & 70.20 & 77.32 & 65.83 & 58.10 & 82.57 \\
\arrayrulecolor{green4}\bottomrule
\end{tabular}
}
\label{tab:results}
\end{table*}

\newpage
\section*{Supplementary Note 2 -- Heat Maps of Error Analysis}
In our analysis, a series of detailed heat maps were utilized to systematically visualize and assess the model’s performance across various demographic intersections. These visualizations are crucial for uncovering demographic-specific biases and identifying areas where the model's predictions may be lacking in accuracy and fairness.
The heat maps, shown in Supplementary Figures \ref{fig:age_race}, \ref{fig:age_gender}, \ref{fig:age_pd}, \ref{fig:race_gender}, \ref{fig:race_pd}, and \ref{fig:gender_pd}, demonstrate how different demographic factors --- age, ethnicity, sex, and PD diagnosis --- affect error rates and coverage. Color codings within these maps highlight the severity of errors, with a stronger blue indicating higher error rates. This provides a clear visual representation of the underperforming demographic groups, offering precise insights for targeted model improvements.
By pinpointing and addressing these specific areas of lagging performance, the heat maps drive our efforts to refine the model, aiming for equitable and accurate outcomes across all demographic groups.

\begin{figure*}[!htbp]
    \begin{subfigure}[t]{1\textwidth}
    \centering
    \includegraphics[width=1\linewidth]{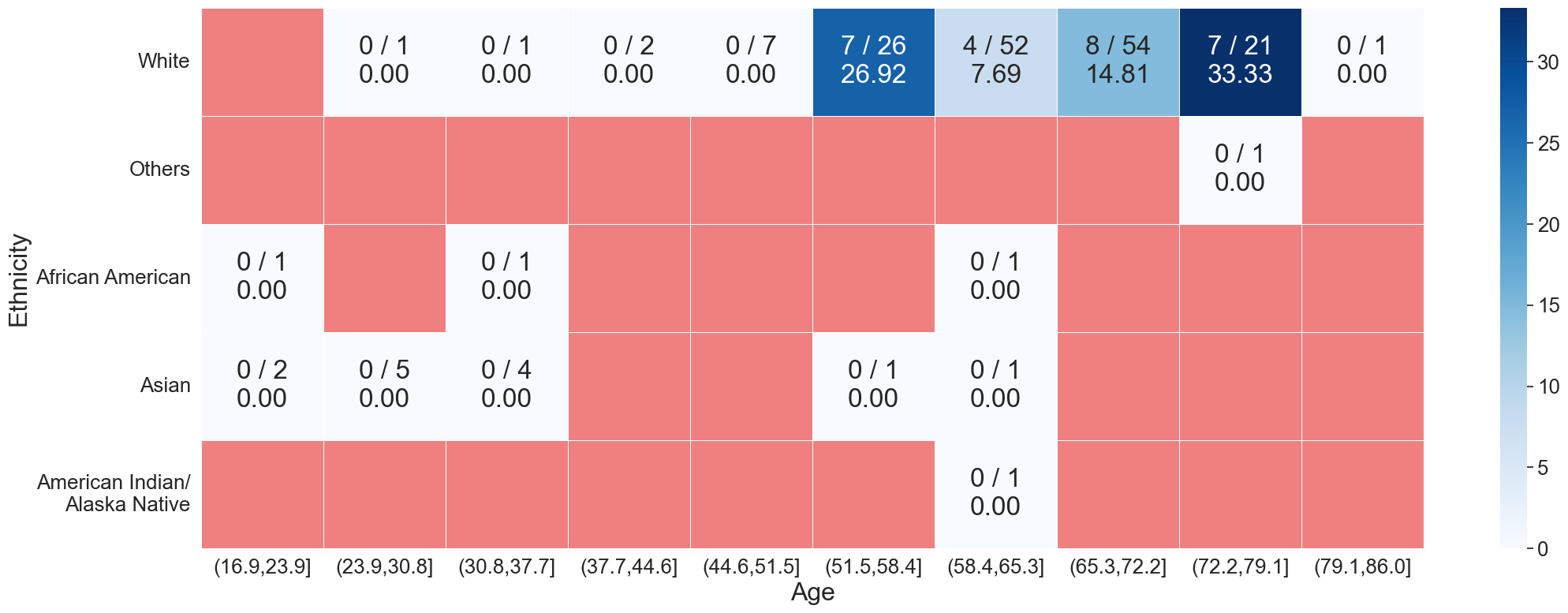}
    \caption{Age vs. Ethnicity}
    \label{fig:age_race}
    \end{subfigure}
    
    \bigskip
    
  \begin{subfigure}[t]{1\textwidth}
    \centering
    \includegraphics[width=1\linewidth]{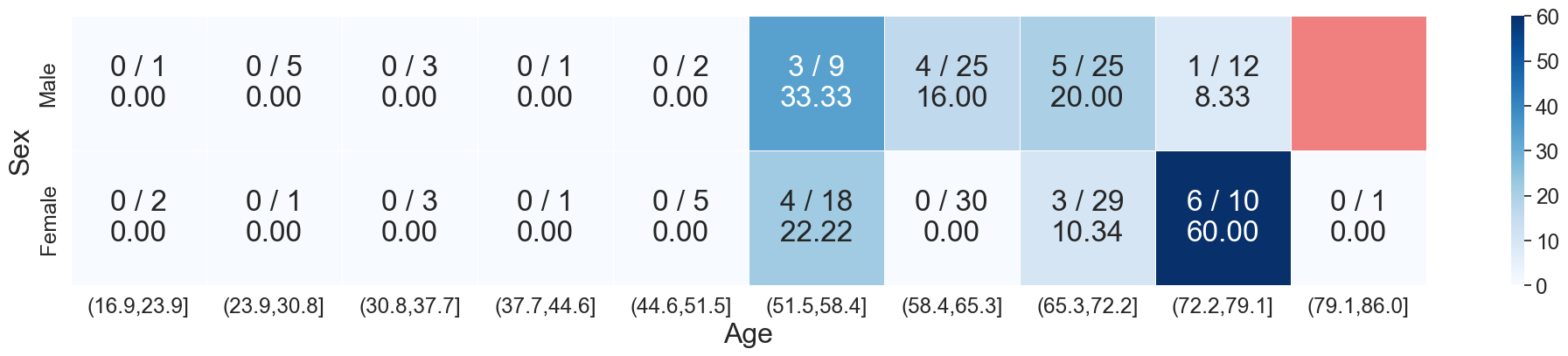}
    \caption{Age vs. Sex}
    \label{fig:age_gender}
  \end{subfigure}
  \bigskip
     \begin{subfigure}[t]{1\textwidth}
    \centering
    \includegraphics[width=1\linewidth]{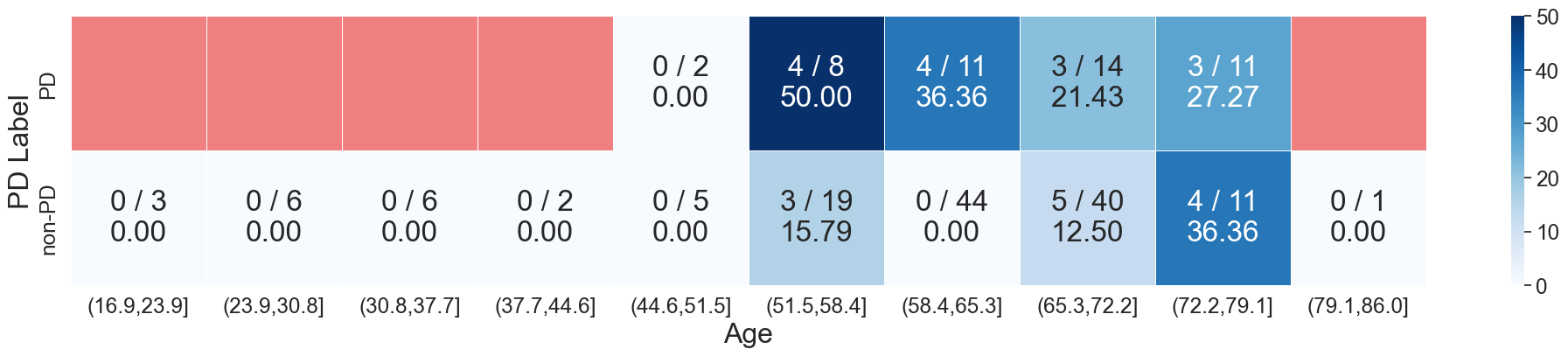}
    \caption{Age vs. PD Label}
    \label{fig:age_pd}
  \end{subfigure} 
  \bigskip
    \centering
    \begin{subfigure}[t]{0.57\textwidth}
    \includegraphics[width=1\linewidth]{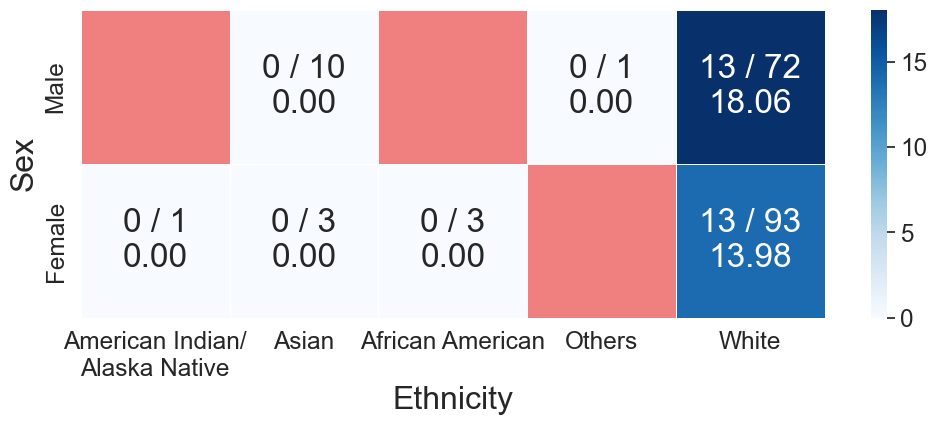}
    \caption{Ethnicity vs. Sex}
    \label{fig:race_gender}
  \end{subfigure}
    \caption{\textbf{Heat maps of error rates and error coverage among demographic cohorts.} A stronger blue color indicates a cohort has a higher error rate, while red indicates that the cohort is empty. Within each cohort: upper row numbers represent the misclassified counts out of the total counts (i.e. incorrect/total); bottom row number represent error rate in percentage.}
  \label{fig:error_heatmap1}
\end{figure*}
\begin{figure*}[!htbp]
     \begin{subfigure}[t]{0.57\textwidth}
    \centering
    \includegraphics[width=1\linewidth]{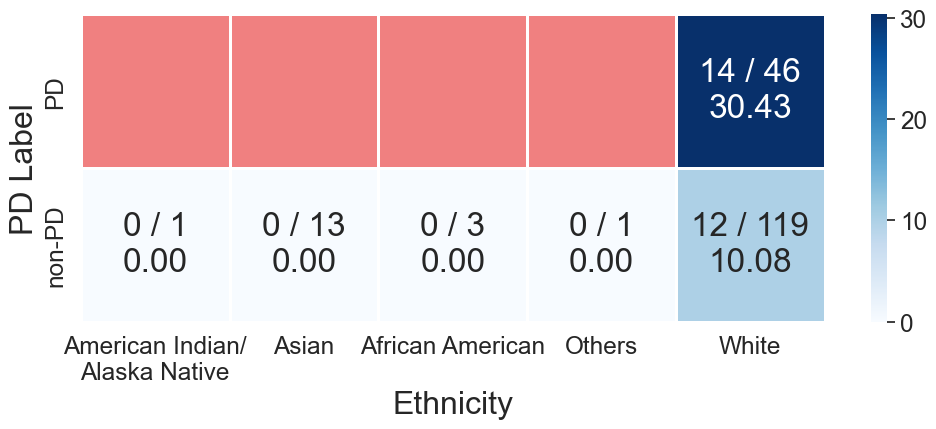}
    \caption{Ethnicity vs. PD Label}
    \label{fig:race_pd}
  \end{subfigure}
    \begin{subfigure}[t]{0.31\textwidth}
    \centering
    \includegraphics[width=1\linewidth]{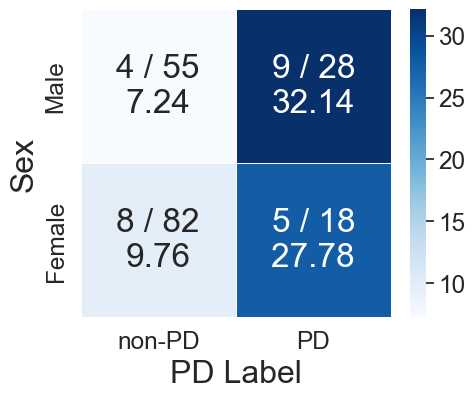}
    \caption{Sex vs. PD Label}
    \label{fig:gender_pd}
  \end{subfigure}
    \caption{\textbf{(Continued from Supplementary Figure 1) Heat maps of error rates and error coverage among demographic cohorts.} A stronger blue color indicates a cohort has a higher error rate, while red indicates that the cohort is empty. Within each cohort: Upper row numbers represent the misclassified counts out of the total counts (i.e. incorrect/total); bottom row number represent error rate in percentage.}
  \label{fig:error_heatmap2}
\end{figure*}

\newpage

\section*{Supplementary Note 3 --- Hyperparameter Tuning}

Supplementary Table~\ref{tab:hyperparameters} demonstrates the hyperparameter space with a quick description. For the hybrid fusion approach, we adapted a similar hyperparameter search space but incorporated slight modifications to account for the added reconstructive loss and the projection method. {Table~\ref{tab:hyperparameters_best} lists the values for hyperparameters for the best performing model.}

\renewcommand{\arraystretch}{1.4}
\begin{table}[!htbp]
\centering
\caption{Hyperparameter Space for Model Optimization}
\begin{tabular}{p{3cm}llp{5.5cm}}
\toprule
\textbf{Parameter} & \textbf{Values (Range)} & \textbf{Distribution} & \textbf{Description} \\ \midrule
Batch Size & \{128, 256, 512, 1024\} & Categorical & Sizes of batches for processing training data. \\ 
Beta1 & 0.9 to 0.99 & Uniform & Exponential decay rate for the first moment estimates in the Adam optimizer. \\ 
Beta2 & 0.99 to 0.9999 & Uniform & Controls the exponential decay rate for the second-moment estimates. \\ 
Correlation Threshold (corr\_thr) & \{0.8, 0.85, 0.9, 0.95\} & Categorical & Threshold to decide when features are considered too correlated. \\ 
Dropping Correlated Features (drop\_correlated) & \{"yes", "no"\} & Categorical & Whether to drop features that exceed the specified correlation threshold. \\ 
Gamma & 0.5 to 0.95 & Uniform & Adjusts the learning rate decay in learning rate schedulers. \\ 
Learning Rate & 0.05 to 0.8 & Uniform & Sets the step size at each iteration while moving toward a minimum of a loss function. \\ 
Minority Oversample & \{"yes", "no"\} & Categorical & Whether to apply oversampling techniques to balance class distribution. \\ 
Model & \{ANN, ShallowANN\} & Categorical & The type of model architecture to be used. \\ 
Momentum & 0.1 to 1 & Uniform & Influences the incorporation of the previous update steps into the current step. \\ 
Number of Epochs (num\_epochs) & 2 to 500 & Integer Uniform & The total number of times that the learning algorithm will work through the entire training dataset. \\ 
Optimizer & \{AdamW, SGD\} & Categorical & The method used to update the weights of the network. \\ 
Patience & 1 to 5 & Integer Uniform & Number of epochs with no improvement after which training will be stopped. \\ 
Random State & 100 to 999 & Integer Uniform & The seed used by the random number generator. \\ 
Scaling Method & \{StandardScaler, MinMaxScaler\} & Categorical & Method for scaling feature data before feeding it into the model. \\ 
Scheduler & \{"step", "reduce"\} & Categorical & Strategy for adjusting the learning rate during training. \\ 
Seed & 100 to 999 & Integer Uniform & Sets the seed for initializing the model, ensuring consistent initial conditions. \\ 
Step Size & 1 to 30 & Integer Uniform & Determines the interval of epochs after which specific actions are taken. \\ 
Use Feature Scaling & \{"yes", "no"\} & Categorical & Whether or not to apply feature scaling. \\ 
Use Scheduler & \{"yes", "no"\} & Categorical & Determines whether a learning rate scheduler is used to optimize training. \\ \bottomrule
\end{tabular}
\label{tab:hyperparameters}
\end{table}

\renewcommand{\arraystretch}{1.2}
\begin{table}[!htbp]
\centering
\caption{{Hyperparameters for the Best Model (WavLM Features Projected to ImageBind Dimension)}}
\begin{tabular}{ll}
\toprule
\textbf{Parameter} & \textbf{Best Value} \\ 
\midrule
Batch Size & 128 \\ 
Beta1 & 0.9084719350261068\\ 
Beta2 & 0.9940871758715644\\ 
Correlation Threshold (corr\_thr) & 0.85\\ 
Dropping Correlated Features (drop\_correlated) & "yes" \\ 
Gamma & 0.6033860204614545 \\ 
Learning Rate & 0.3674643450313223 \\ 
Minority Oversample & "no" \\ 
Model & ANN\\ 
Momentum & 0.8075456327084843 \\ 
Number of Epochs (num\_epochs) & 86 \\ 
Optimizer & SGD \\ 
Patience & 5 \\ 
Random State & 621 \\ 
Scaling Method & MinMaxScaler\\ 
Scheduler & "reduce"\\ 
Seed & 191\\ 
Step Size & 17 \\ 
Use Feature Scaling & "yes" \\ 
Use Scheduler &  "no"\\ 
Weight of Cosine Loss & 68 \\ 
Weight of Prediction Loss & 87 \\ 
Weight of Reconstruction Loss &  48\\ 
\bottomrule
\end{tabular}
\label{tab:hyperparameters_best}
\end{table}

\newpage

\section*{{Supplementary Note 4 --- Statistical Relevance Test of Classical Acoustic Features}}

\begin{table*}[!htbp]
\caption{{Statistical relevance test of the classical acoustic features to differentiate PD and Non-PD} }
\resizebox{1.0\linewidth}{!}{%
\begin{tabular}{llccc}
\toprule
\textbf{Extracted Feature} & \textbf{Related Speech Feature} & \textbf{Mann-Whitney Statistic} & \textbf{P-value} & \textbf{Statistical Significance}\\
\midrule

cepm2 & MFCC 2 & $3.96 \times 10^{5}$ & $3.78 \times 10^{-28}$ & Yes\\
relbandpower3 & Relative Band Power (High Frequency) & $1.72 \times 10^{5}$ & $2.12 \times 10^{-25}$ & Yes\\
cepm4 & MFCC 4 & $3.84 \times 10^{5}$ & $1.07 \times 10^{-21}$ & Yes\\
cepj0 & Derived MFCC 0 (Jitter-related) & $2.12 \times 10^{5}$ & $1.80 \times 10^{-21}$ & Yes\\
relbandpower1 & Relative Band Power (Mid-Low Frequency) & $1.83 \times 10^{5}$ & $2.43 \times 10^{-19}$ & Yes\\
ppe & Pitch Period Entropy (PPE) & $2.22 \times 10^{5}$ & $6.04 \times 10^{-17}$ & Yes\\
relbandpower2 & Relative Band Power (Mid-High Frequency) & $1.91 \times 10^{5}$ & $9.21 \times 10^{-16}$ & Yes\\
cepj3 & Derived MFCC 3 (Jitter-related) & $2.36 \times 10^{5}$ & $5.42 \times 10^{-12}$ & Yes\\
cepj4 & Derived MFCC 4 (Jitter-related) & $2.38 \times 10^{5}$ & $3.06 \times 10^{-11}$ & Yes\\
cepj5 & Derived MFCC 5 (Jitter-related) & $2.39 \times 10^{5}$ & $7.00 \times 10^{-11}$ & Yes\\
cepj1 & Derived MFCC 1 (Jitter-related) & $2.40 \times 10^{5}$ & $1.17 \times 10^{-10}$ & Yes\\
ashr & Relative Shimmer & $2.41 \times 10^{5}$ & $3.71 \times 10^{-10}$ & Yes\\
ash & Shimmer (Amplitude) & $2.41 \times 10^{5}$ & $3.80 \times 10^{-10}$ & Yes\\
cepm5 & MFCC 5 & $3.53 \times 10^{5}$ & $6.80 \times 10^{-10}$ & Yes\\
cepm1 & MFCC 1 & $3.53 \times 10^{5}$ & $7.86 \times 10^{-10}$ & Yes\\
cepj6 & Derived MFCC 6 (Jitter-related) & $2.46 \times 10^{5}$ & $9.77 \times 10^{-9}$ & Yes\\
cepj2 & Derived MFCC 2 (Jitter-related) & $2.46 \times 10^{5}$ & $1.18 \times 10^{-8}$ & Yes\\
cepm0 & MFCC 0 & $2.46 \times 10^{5}$ & $1.37 \times 10^{-8}$ & Yes\\
cepm9 & MFCC 9 & $2.48 \times 10^{5}$ & $3.93 \times 10^{-8}$ & Yes\\
alpha & Alpha Ratio & $3.41 \times 10^{5}$ & $1.12 \times 10^{-6}$ & Yes\\
f0m & Pitch (F0) & $2.58 \times 10^{5}$ & $1.06 \times 10^{-5}$ & Yes\\
cepj8 & Derived MFCC 8 (Jitter-related) & $2.64 \times 10^{5}$ & $1.60 \times 10^{-4}$ & Yes\\
f0jr & Relative Jitter & $2.67 \times 10^{5}$ & $6.16 \times 10^{-4}$ & Yes\\
f0j & Jitter & $2.67 \times 10^{5}$ & $7.13 \times 10^{-4}$ & Yes\\
cepm12 & MFCC 12 & $2.69 \times 10^{5}$ & $0.0014$ & Yes\\
cepm7 & MFCC 7 & $2.71 \times 10^{5}$ & $0.0037$ & Yes\\
cepj11 & Derived MFCC 11 (Jitter-related) & $2.75 \times 10^{5}$ & $0.0114$ & Yes\\
cepj10 & Derived MFCC 10 (Jitter-related) & $2.79 \times 10^{5}$ & $0.0353$ & Yes\\
cepm6 & MFCC 6 & $3.16 \times 10^{5}$ & $0.0422$ & Yes\\
relbandpower0 & Relative Band Power (Low Frequency) & $2.59 \times 10^{5}$ & $0.0470$ & Yes\\
cepm8 & MFCC 8 & $3.14 \times 10^{5}$ & $0.0606$ & No\\
Hnorm & Harmonic Normalization (Hnorm) & $2.82 \times 10^{5}$ & $0.0825$ & No\\
cepj9 & Derived MFCC 9 (Jitter-related) & $2.84 \times 10^{5}$ & $0.1256$ & No\\
cepm11 & MFCC 11 & $3.10 \times 10^{5}$ & $0.1734$ & No\\
cepj7 & Derived MFCC 7 (Jitter-related) & $2.87 \times 10^{5}$ & $0.2284$ & No\\
cepm10 & MFCC 10 & $2.88 \times 10^{5}$ & $0.2778$ & No\\
cepj12 & Derived MFCC 12 (Jitter-related) & $2.96 \times 10^{5}$ & $0.8364$ & No\\
cepm3 & MFCC 3 & $2.98 \times 10^{5}$ & $0.9397$ & No\\

\bottomrule
\end{tabular}
}
\end{table*}

\end{document}